\documentclass[fleqn,usenatbib]{mnras}

\usepackage{mathptmx}

\usepackage[T1]{fontenc}

\usepackage[normalem]{ulem} 

\DeclareRobustCommand{\VAN}[3]{#2}
\let\VANthebibliography\thebibliography
\def\thebibliography{\DeclareRobustCommand{\VAN}[3]{##3}\VANthebibliography}

\usepackage{graphicx}	
\usepackage{amsmath}	
\usepackage{amssymb}	
\usepackage{color}
\usepackage{booktabs}
\usepackage{hyperref}
\usepackage{empheq}

\usepackage{graphicx}	
\usepackage{amsmath}	
\usepackage{amssymb}	
\usepackage{color}
\usepackage{booktabs}
\usepackage{hyperref}
\usepackage{empheq}

\mathchardef\mhyphen="2D

\newcommand{\di}{\mathrm{d}}
\newcommand{\bfx}{\mathbf{x}}

\newcommand{\degree}{\ensuremath{^\circ}}

\newcommand{\pc}{\,{\rm pc}}
\newcommand{\kpc}{\,{\rm kpc}}
\newcommand{\Myr}{\,{\rm Myr}}
\newcommand{\Gyr}{\,{\rm Gyr}}
\newcommand{\kms}{\,{\rm km\, s^{-1}}}

\newcommand{\pa}{\partial}
\newcommand{\cmthree}{\,{\rm cm^{-3}}}

\newcommand{\Msun}{\,{\rm M_\odot }}
\newcommand{\Msunyr}{\,{\rm M_\odot \, yr^{-1}}}

\title[Star formation in the CMZ]{Simulations of the Milky Way's central molecular zone - II. Star formation}
\author[Sormani et al.]{Mattia C. Sormani$^{1}$, Robin G. Tress$^1$, Simon C.O. Glover$^1$, Ralf S. Klessen$^{1,2}$,  \newauthor Cara D. Battersby$^3$, Paul C. Clark$^4$, H Perry Hatchfield$^3$ and Rowan J. Smith$^5$\\
$^1$Universit\"{a}t Heidelberg, Zentrum f\"{u}r Astronomie, Institut f\"{u}r theoretische Astrophysik, Albert-Ueberle-Str. 2, 69120 Heidelberg, Germany \\
$^2$Universit\"at Heidelberg, Interdiszipli\"ares Zentrum f\"ur Wissenschaftliches Rechnen, Im Neuenheimer Feld 205, 69120 Heidelberg, Germany \\
$^3$University of Connecticut, Department of Physics, 196 Auditorium Road, Unit 3046, Storrs, CT 06269, USA\\
$^4$School of Physics and Astronomy, Queen's Buildings, The Parade, Cardiff University, Cardiff, CF24 3AA, UK \\
$^5$Jodrell Bank Centre for Astrophysics, School of Physics and Astronomy, University of Manchester, Oxford Road, Manchester M13 9PL, UK \\
}

\begin{document}

\date{} 
\maketitle

\begin{abstract}
The Milky Way's central molecular zone (CMZ) has emerged in recent years as a unique laboratory for the study of star formation. Here we use the simulations presented in Tress et al. 2020 to investigate star formation in the CMZ. These simulations resolve the structure of the interstellar medium at sub-parsec resolution while also including the large-scale flow in which the CMZ is embedded. Our main findings are as follows. (1) While most of the star formation happens in the CMZ ring at $R\gtrsim100 \pc$, a significant amount also occurs closer to SgrA* at $R \lesssim 10\pc$. (2) Most of the star formation in the CMZ happens downstream of the apocentres, consistent with the ``pearls-on-a-string'' scenario, and in contrast to the notion that an absolute evolutionary timeline of star formation is triggered by pericentre passage. (3) Within the timescale of our simulations ($\sim100$~Myr), the depletion time of the CMZ is constant within a factor of $\sim2$. This suggests that variations in the star formation rate are primarily driven by variations in the mass of the CMZ, caused for example by AGN feedback or externally-induced changes in the bar-driven inflow rate, and not by variations in the depletion time. (4) We study the trajectories of newly born stars in our simulations. We find several examples that have age and 3D velocity compatible with those of the Arches and Quintuplet clusters. Our simulations suggest that these prominent clusters originated near the collision sites where the bar-driven inflow accretes onto the CMZ, at symmetrical locations with respect to the Galactic centre, and that they have already decoupled from the gas in which they were born.
\end{abstract}

\begin{keywords}
Galaxy: centre - Galaxy: kinematics and dynamics - ISM: kinematics and dynamics - ISM: clouds - ISM: evolution - stars: formation
\end{keywords}

\section{Introduction} \label{sec:intro}

The central molecular zone (CMZ, $R\lesssim200\pc$) is the Milky Way's counterpart of the star-forming nuclear rings that are commonly found in the central regions of external barred galaxies such as NGC 1300 (see for example the atlas of nuclear rings of \citealt{Comeron+2010}). Being a hundred times closer than the nucleus of the next comparable galaxy,  Andromeda, it offers us the possibility to study a nuclear ring in unique detail.

The CMZ has emerged in the last decade as a unique laboratory for the study of star formation \citep[e.g.][]{Molinari+2011,Kruijssen+14b,Armillotta+2019}. The main reason is that the environmental conditions in which stars are born are more extreme than anywhere else in the Galaxy. Indeed, the physical properties of the interstellar medium (ISM) in the CMZ are substantially different from those in the Galactic disc: average gas volume densities \citep{GuestenHenkel1983,Walmsley+1986,Longmore+2017,Mills+2018}, temperatures \citep{Immer+2016,Ginsburg+2016,Krieger+2017,Oka+2019}, velocity dispersions \citep{Shetty+2012,Federrath+2016}, and magnetic field strengths \citep{Morris2015,Mangilli+2019} are all much higher than in the disc. The interstellar radiation field and higher cosmic ray ionisation rate \citep{Clark+2013,Ginsburg+2016,Oka+2019} are also much stronger. In addition, the CMZ region is  characterised by the presence of Galactic outflows \citep{Ponti+2019}, by the widespread presence of radio-emitting magnetised non-thermal filaments \citep{Heywood+2019}, and by a strong hydrodynamical interaction with the larger-scale gas inflow driven by the Galactic bar \citep{Sormani+2018a}. The star formation process, which is determined by the complex interplay of all these physical agents, is therefore expected to proceed differently in the CMZ. Observations confirm this, by showing that the CMZ does not obey some star formation relations that are valid in the disc \citep{Longmore+2013a,Kauffmann+2017b}. Hence, understanding star formation in the CMZ is important for understanding the star formation process in extreme environments, as well as in general by probing a peculiar corner of parameter space. 

In a companion paper (\citealt{Tress+2020}, hereafter Paper I) we have presented sub-parsec resolution hydrodynamical simulations and have used them to study the gas dynamics in the CMZ. In this paper, we use the same simulations to investigate star formation.

Open questions that we address in the current work include: 
\begin{enumerate}
\item What is the temporal distribution of star formation in the CMZ? (Section~\ref{sec:temporalsf})
\item What is spatial distribution of star formation in the CMZ? (Section~\ref{sec:spatialsf})
\item What is the impact of the orbital dynamics on star formation? Can we identify an absolute evolutionary timeline of star formation as suggested by \cite{Longmore+2013b} and \cite{Kruijssen+2015}? (Section~\ref{sec:timelinesf})
\item What drives the time variability of star formation in the CMZ? (Section~\ref{sec:sfrate})
\item Are the Arches and Quintuplet cluster on a common orbit with gas in the CMZ ring \citep{Kruijssen+2015} or are they on other types of orbits \citep{Stolte+2008}? (Section~\ref{sec:arches})
\end{enumerate}

The paper is structured as follows. In Section~\ref{sec:methods} we give a brief summary of our numerical simulations. In Section~\ref{sec:sf} we study the temporal and spatial distribution of star formation and the trajectories of newly born stars. In Section~\ref{sec:discussion} we discuss the implications of our results for some of the open questions raised above. We sum up in Section~\ref{sec:conclusion}.

\section{Numerical methods} \label{sec:methods}

Our simulations have been presented in detail in Paper I. Hence we give here only a very brief overview, and refer to that paper for more details.

\subsection{Overview} \label{sec:methods:overview}

The simulations are similar to those we previously discussed in \cite{Sormani+2018a,Sormani+2019}, with the following differences: (i) inclusion of gas self-gravity; and (ii) inclusion of a sub-grid prescription for star formation and stellar feedback. In particular, we employ exactly the same externally-imposed rotating barred potential, the same chemical/thermal treatment of the gas, and the same initial conditions as in \cite{Sormani+2019}. 

We use the moving-mesh code {\sc arepo} \citep{Springel2010,Weinberger+2020}. The simulations are three-dimensional and unmagnetised, and include a live chemical network that keeps track of hydrogen and carbon chemistry. The simulations comprise interstellar gas in the whole inner disc ($R \leq 5 \kpc$) of the Milky Way, which allows us to understand the CMZ in the context of the larger-scale flow, which is important since the CMZ strongly interacts with its surrounding through the bar inflow \citep{Sormani+2018a}. The gas is assumed to flow in a multi-component external rotating barred potential $\Phi_{\rm ext}(\bfx,t)$ which is constructed to fit the properties of the Milky Way. The bar component rotates with a pattern speed $\Omega_{\rm p} = 40\kms \kpc^{-1}$, consistent with the most recent determinations \citep[e.g.][]{SBM2015c,Portail+2017,Sanders+2019}, which places the (only) inner Lindblad resonance (ILR) calculated in the epicyclic approximation at $R_{\rm ILR}=1.1\kpc$ and the corotation resonance at $R_{\rm CR}=5.9\kpc$. The potential is identical to that used in \cite{Sormani+2019} and is described in more detail in the appendix of that paper.

Gas self-gravity is included. The process of star formation and the consequent stellar feedback are modelled as follows (see Section 2 of Paper I for more details):
\begin{enumerate}
\item Gravitationally-collapsing gas that exceeds a density threshold $\rho_{\rm c} = 10^{-20} \: {\rm g \, cm^{-3}}$ is removed from the simulation and replaced with a non-gaseous sink particle, provided that it is unambiguously gravitationally bound and not within the accretion radius of an existing sink particle. The sink particle does not represent an individual star, but rather a small cluster which contains both gas and stars.
\item Once a sink is created, a stellar population is assigned to it by drawing from an initial mass function (IMF) according to the Poisson stochastic method described in \cite{Sormani+2017b}.
\item Sink particles are allowed to accrete mass at later times, provided that the gas is within the sink accretion radius $r_{\rm acc} = 1$~pc and is gravitationally bound to the sink. The stellar population associated with a given sink is updated every time that that sink accretes additional mass.
\item For each massive star ($M \geq 8 \Msun$) assigned to the sink, we produce a supernovae (SNe) event with a time delay which depends on the stellar mass. Each SNe event injects energy and/or momentum into the ISM and gives back to the environment part of the gas ``locked-up'' in the sink. Energy is injected only if the local resolution of the Voronoi mesh is high enough to resolve the supernova remnant at the end of its Sedov-Taylor phase; otherwise, an appropriate amount of momentum is injected instead. SNe feedback is the only type of feedback included in the simulation.
\item When all the SNe associated with a sink have exploded and all of its gas content has been given back to the environment, the sink is converted into a collisionless N-body star particle with a mass equal to the stellar mass of the sink. This N-body particle continues to exist indefinitely in the simulation and affects it through its gravitational potential, but, unlike a sink, it can no longer accrete new gas or form new stars. 
\end{enumerate}

When making projections onto the plane of the Sky, we assume an angle between the Sun-Galactic centre line and the bar major axis of $\phi = 20^\circ$, a Sun-Galactic centre distance of $8.2\kpc$ \citep{Reid+2019,Gravity2019}, and that the Sun is on a circular orbit at $v_\odot = 235\kms$ \citep{Schoenrich+2010,Reid+2019}, as in Paper I.

\subsection{Subdivision in three regions: CMZ, DLR, disc} \label{sec:regions}

\begin{figure}
	\includegraphics[width=\columnwidth]{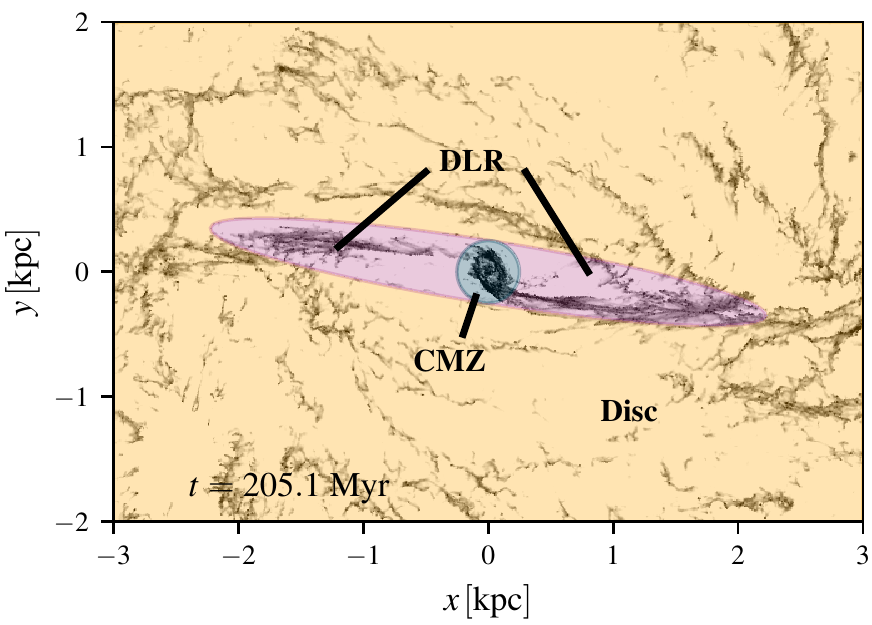}
    \caption{Definition of the three regions (CMZ, DLR, disc) into which we subdivide our simulated Galaxy for subsequent analysis. See Section~\ref{sec:regions} for more details.}
    \label{fig:regions}
\end{figure}

As in Paper I, we subdivide our simulation into three spatial regions in order to facilitate the subsequent analysis (see Figure~ \ref{fig:regions}): 
\begin{itemize}
\item The \emph{CMZ} is defined as the region within cylindrical radius $R \leq 250 \pc$. 
\item The \emph{dust lane region (DLR)} is the elongated transition region between the CMZ and the Galactic disc, where highly non-circular gas motions caused by the bar are present.
\item The \emph{disc} is defined as everything outside the DLR.
\end{itemize}

\section{Star formation} \label{sec:sf}

\subsection{Temporal distribution of star formation} \label{sec:temporalsf}
 
\begin{figure*}
	\includegraphics[width=\textwidth]{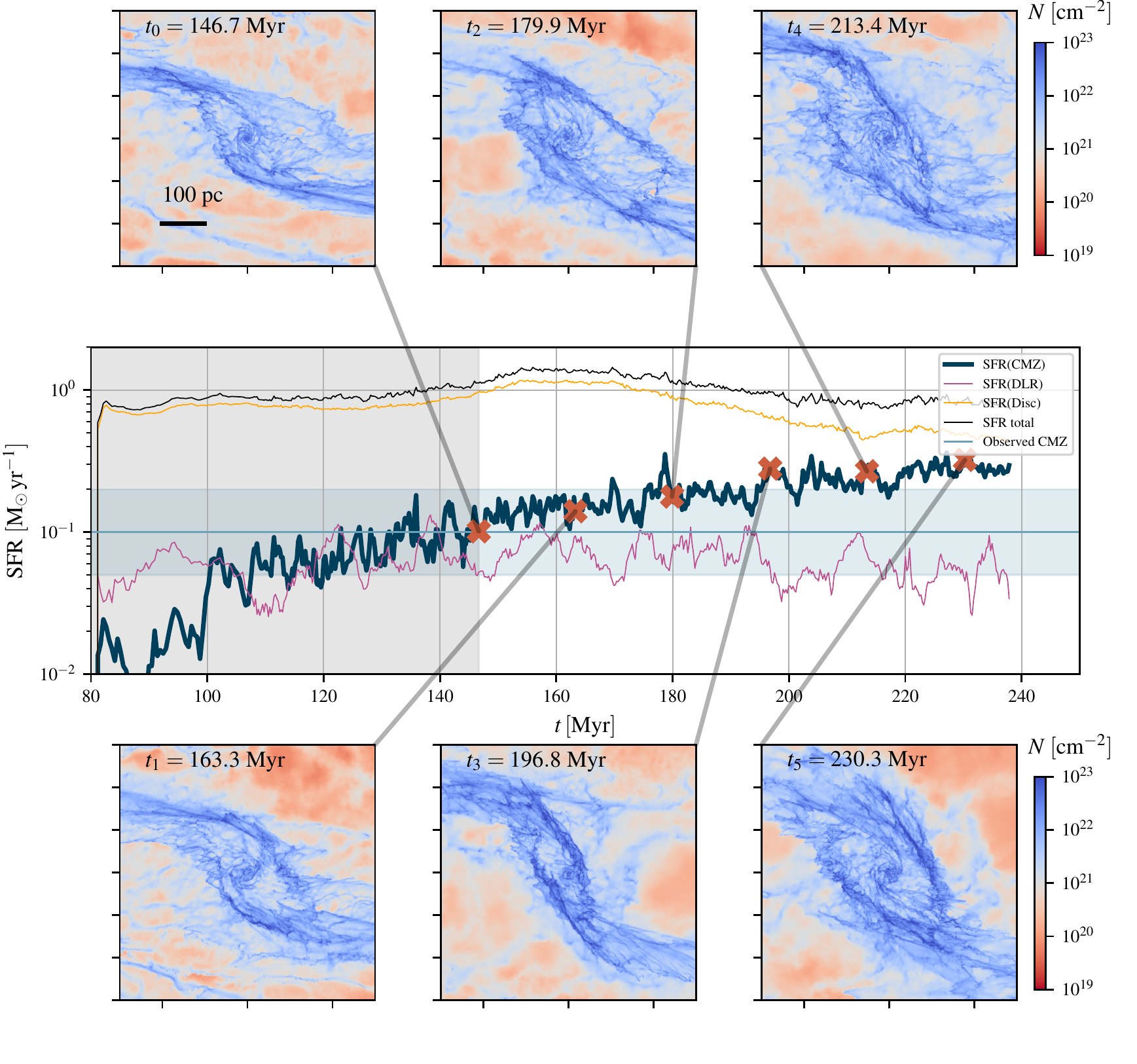}
    \caption{Star formation rate as a function of time in our simulation. The thick blue, thin pink and thin yellow lines are the SFR in the three different spatial regions (CMZ, DLR, disc) in which we have subdivided our simulation (see Figure~\ref{fig:regions}). The thin black line is the total SFR (CMZ+DLR+disc). The insert panels show total gas surface density maps that allow us to correlate the SFR with the instantaneous CMZ morphology. The blue shaded horizontal region indicates the observed current SFR of the CMZ, taken to be in the conservative range $0.05 \mhyphen 0.2 \Msunyr$ (see references in Section \ref{sec:sfrate}) The grey shaded area indicates the times when the bar potential is still gradually turning on (see Section~2.7 in Paper I), which are excluded from the analysis.}
    \label{fig:sfmain}
\end{figure*}

Figure \ref{fig:sfmain} shows the star formation rate (SFR) as a function of time in our simulation, calculated as a running average over the last $0.5 \Myr$. This corresponds to twice the timestep between consecutive simulation outputs, and has been chosen because we want to study where and when star formation is being triggered.\footnote{Observationally determined rates are more often averaged over longer timescales ($\sim 10\Myr$). We will briefly discuss the distribution of older stars in Section \ref{sec:spatialsf}, while we defer a more observationally oriented approach and synthetic observations to future work.} The thin blue line shows the total SFR in the entire simulation box (CMZ + DLR + disc). This is roughly constant at a value of approximately $\sim 1 \Msunyr$, consistent with typically reported values of the MW total SFR derived from observations ($\sim 2 \mhyphen 3 \Msunyr$, e.g. \citealt{KennicuttEvans2012}) when we take into account that our simulated disc only extends to $R \simeq 5 \kpc$, so the total gas mass in the simulation ($\simeq 1.5 \times 10^9 \Msun$) is only $\sim 1/3$ of the total estimated mass in the MW gas disc.

The thick blue line shows the total SFR of the CMZ (defined as the region $R \leq 250 \pc$, see Figure~\ref{fig:regions}). The insert panels correlate the SFR with the CMZ gas morphology at different times. At $t=146 \Myr$ (when the bar potential is fully turned on, see Section~2.7 in Paper I), the SFR in the CMZ has a value of $\sim 0.1 \Msunyr$, consistent with observational estimates (see \citealt{YusefZadeh+2009,Immer+2012,Longmore+2013a,Barnes+2017} and Section~\ref{sec:sfrate}), and the total gas mass of the CMZ is $\sim 4 \times 10^7 \Msun$, which also agrees well with observational values ($\sim 5 \times 10^7 \Msun$, \citealt{Dahmen+1998,Longmore+2013a}). 

\begin{figure}
	\includegraphics[width=\columnwidth]{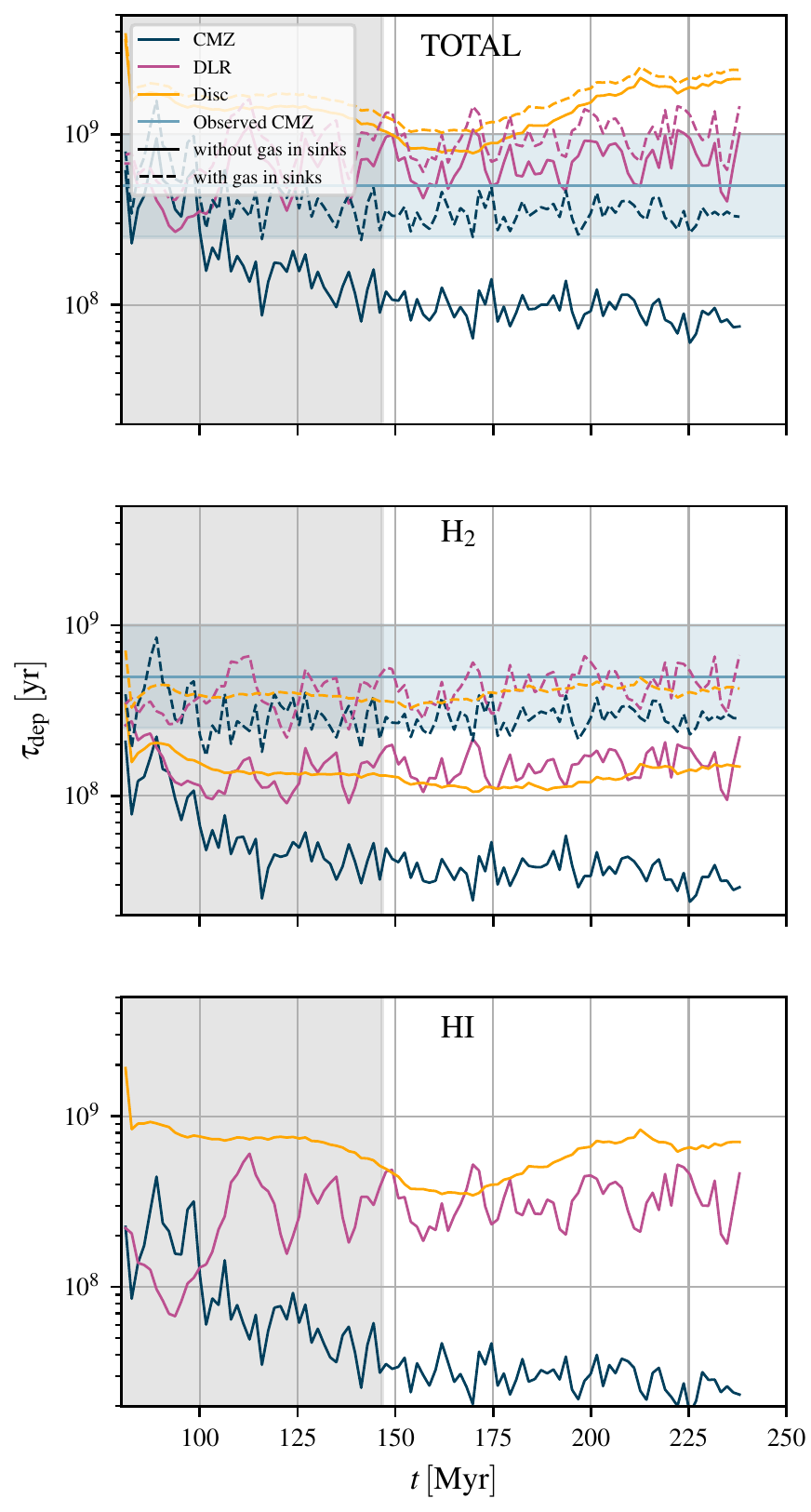}
    \caption{Depletion time ($\tau_{\rm dep}$) as a function of simulation time ($t$) for the various regions defined in Figure~\ref{fig:regions}.  The blue shaded region indicates the observed depletion time of the CMZ, $0.25-1 \Gyr$, obtained by dividing the estimated total molecular mass of the CMZ ($5 \times 10^7 \Msun$, see references in Section \ref{sec:temporalsf}) by the observed SFR of the CMZ shown in Figure \ref{fig:sfmain} ($0.05 \mhyphen 0.2 \Msunyr$). The grey shaded area indicates where the bar potential is gradually turning on, which are excluded from the analysis.}
    \label{fig:tdepl}
\end{figure}

At later times ($t>146 \Myr$) the SFR of the CMZ slowly but steadily increases with time, with small fluctuations on short timescales ($\sim 1 \Myr$) of a factor of $\sim 2 \mhyphen 3$. This increase in the SFR is mirrored by an increase in the total gas mass of the CMZ (see blue lines in Figure~19 of Paper I). The depletion time, which is defined as the ratio between the mass and the SFR ($\tau_{\rm depl} = M / {\rm SFR}$), is shown by the blue dashed line in Figure~\ref{fig:tdepl}. It is approximately constant in time. Therefore, the SFR in the CMZ in our simulation is roughly proportional to its total mass, and variations in the value of the SFR are determined by variations in the total mass. 

Figure \ref{fig:tdepl} also shows that the depletion time in the disc (yellow dashed line) is a factor of $\sim 5$ higher than the depletion time in the CMZ (blue dashed line). Therefore, while the depletion time of each region is approximately constant in time, there are spatial variations when considering different portions of the Galaxy. The variations in the depletion times can be explained by the different stellar gravitational potential, whose vertical gradient is stronger in the CMZ than in the disc. This can be seen as follows. For a medium in which the turbulence is driven by supernovae feedback and assuming that the vertical force of the gravitational potential is balanced by the turbulent pressure (both conditions that are approximately verified in our simulations) the analytical model of \cite{OstrikerShetty2011} predicts that (see their Equation 13):
\begin{equation} \label{eq:ostrikershetty}
\Sigma_{\rm SFR} \propto (1 + \chi) \Sigma^2\,,
\end{equation}
where $\Sigma_{\rm SFR}$ is the SFR surface density, $\Sigma$ is the total gas surface density, $\chi=2C/(1 + \sqrt{1+4C})$, $C= 8 \zeta_{\rm d} \rho_{\rm b} \sigma_z^2/(3\pi G \Sigma^2)$, $\sigma_{\rm z}$ is the vertical velocity dispersion, $\rho_{b}$ is the stellar midplane density (which is proportional to the strength of the gravitational potential), $G$ is the gravitational constant and $\zeta_{\rm d} \simeq 1/3$ is a numerical factor (unimportant here).  In the limit that the gravitational potential of the stars dominates over the gravitational potential of the gas disc, as is the case for the present simulations, we have $C\gg 1$, $\chi \simeq (2C)^{1/2}$ and therefore $\Sigma_{\rm SFR} \propto \rho_b^{1/2} \Sigma$. The depletion time is then $\tau_{\rm depl} = \Sigma/\Sigma_{\rm SFR} \propto \rho_b^{-1/2}$. For the potential employed in our simulations, we find $[\rho_b(R=150\pc)/\rho_b(R=3\kpc)]^{-1/2} \simeq 6$, in good agreement with the results in Figure \ref{fig:tdepl} considering the uncertainties present both in the simulations and in the simplifying assumptions on which the theory of \cite{OstrikerShetty2011} is based. The agreement between our simulation and the theory of \cite{OstrikerShetty2011} is consistent with the fact that the integrated properties of the CMZ follow well star formation relations based on the total or molecular gas surface density, such as the Schmidt-Kennicutt or the \cite{Bigiel+2008} relation, and only become peculiar when considering the very dense gas (see Section \ref{sec:sfrelations}).

Another factor that is likely to contribute to lowering the depletion time in the CMZ, and which is not accounted for in the vertical equilibrium theories of \cite{Ostriker+2010} and \cite{OstrikerShetty2011}, is the increased number of shocks due to the large-scale bar flow, which cause local compressions and therefore enhanced star formation \citep{MacLowKlessen2004,KlessenGlover2016}.

How would the CMZ mass/SFR evolve if we continue our simulation beyond the maximum time shown in Figure \ref{fig:sfmain}? Assuming that the depletion time remains constant at the value $\tau_{\rm depl, CMZ} \simeq 4 \times 10^8 \, {\rm yr}$ inferred from Figure \ref{fig:tdepl}, we might extrapolate that the mass of the CMZ would keep increasing until the SFR matches the bar-driven inflow rate. For an inflow rate of $\dot{M} \simeq 1 \Msunyr$ (see Paper I), the equilibrium CMZ mass would be $ \dot{M} \tau_{\rm depl, CMZ} \simeq 4\times10^8 \Msun$. However, there are several factors that might invalidate this extrapolation: (i) at a mass $\simeq 4 \times 10^8 \Msun$, the gravitational potential of the gas would become comparable to that of the stars, which would affect the depletion time (see discussion immediately after Equation \ref{eq:ostrikershetty} above); (ii) at a SFR of $\simeq 1 \, \Msunyr$,  the increased SN feedback rate might also change the depletion time; (iii) the bar-driven inflow rate will decrease once the reservoir at $R\gtrsim 3\kpc$ gets depleted as the simulation progresses. In the real Galaxy, additional processes not included in our simulation such as expulsion of gas due to AGN feedback, Galactic winds and externally-driven variations in the bar-driven inflow rate are also likely to modify the mass of the CMZ on comparable or even shorter timescales (see also the discussion in Section \ref{sec:sfrate}).

\subsection{Spatial distribution of star formation} \label{sec:spatialsf}

\begin{figure*}
	\includegraphics[width=\textwidth]{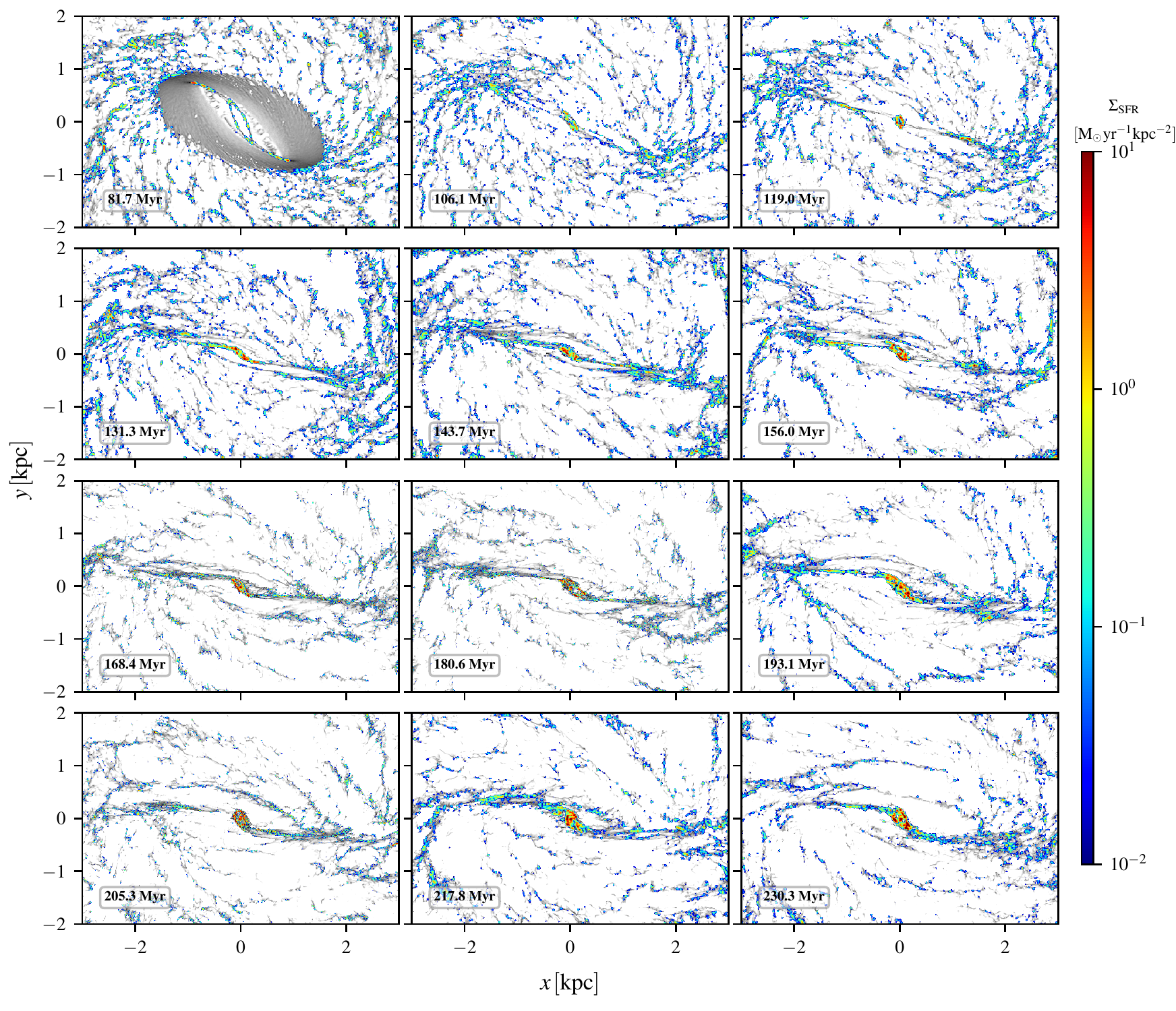}
    \caption{SFR density for various snapshots in our simulation. Shown is the very recent ($0.5 \Myr$) star formation. The grey background shows the H$_2$ surface density. Compare with the time-averaged SFR density shown in the bottom panel of Figure~\ref{fig:average}.}
    \label{fig:SigmaSFR_v_time}
\end{figure*}

\begin{figure*}
	\includegraphics[width=\textwidth]{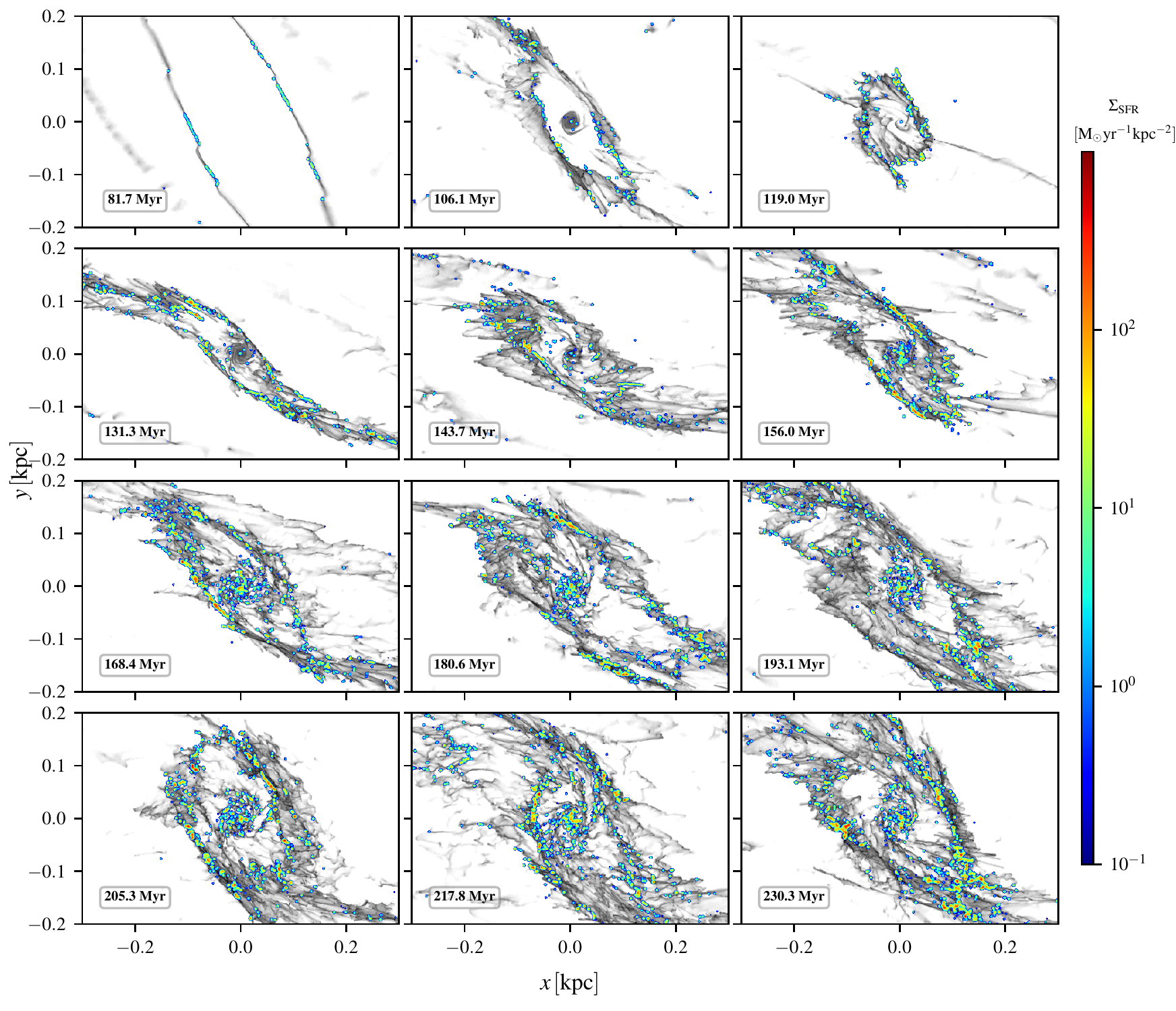}
    \caption{Same as Figure~\ref{fig:SigmaSFR_v_time}, but zooming onto the CMZ. Compare with the time-averaged SFR density shown in the bottom panel of Figure~\ref{fig:average}.}
    \label{fig:SigmaSFR_v_time_CMZ}
\end{figure*}

Figures \ref{fig:SigmaSFR_v_time} and \ref{fig:SigmaSFR_v_time_CMZ} show the spatial distribution of the SFR density in a typical simulation snapshot. As before, the SFR is calculated as the running average over the last $0.5 \Myr$. As expected, star formation occurs predominantly where gas is densest. 

\begin{figure}
	\includegraphics[width=\columnwidth]{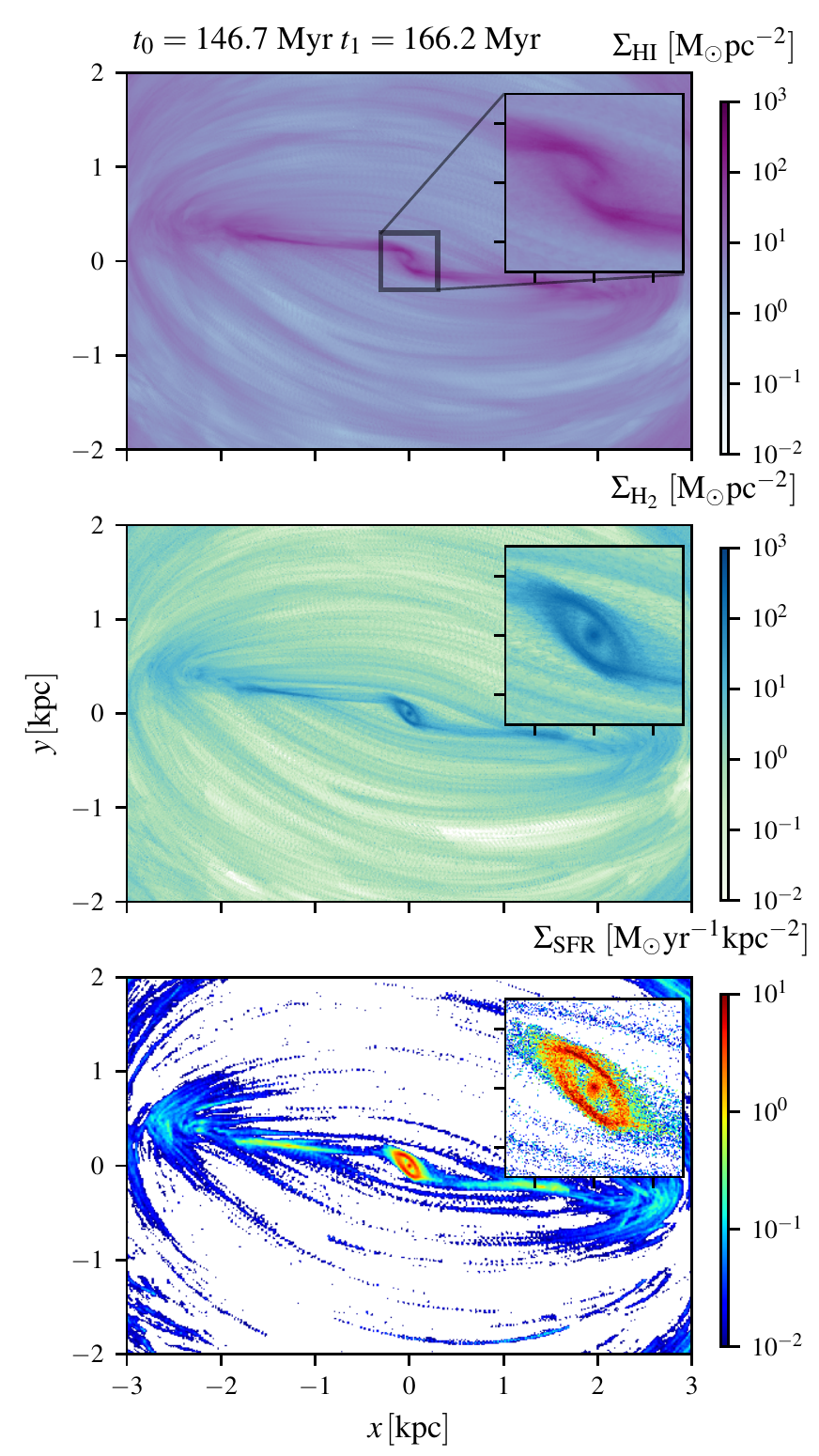}
    \caption{Time-averaged plot of \emph{top:} H$_2$ surface density; \emph{middle}: H{\sc i} surface density; \emph{bottom}: star formation rate (SFR) density. The average is calculated over the range $t = 146.7 \mhyphen 166.2 \Myr$. The ``stripes'' in the SFR rate originate from individual molecular clouds that form stars while following $x_1$ orbits.}
    \label{fig:average}
\end{figure}

It is instructive to compare the ``instantaneous'' SFR density (Figures~\ref{fig:SigmaSFR_v_time} and \ref{fig:SigmaSFR_v_time_CMZ}) with the time-averaged SFR density (bottom panel in Figure~\ref{fig:average}). This comparison shows very clearly that while the time-averaged distribution is smooth, the instantaneous SFR density can have complex and transient morphologies which deviate significantly from the averaged morphology. In particular, the time-averaged star formation in the CMZ is smoothly distributed along an elliptical ring, while looking at the instantaneous SFR does not always give the impression of a ring. The size of the ring is significantly smaller than the ILR calculated in the epicyclic approximation, consistent with previous studies \citep[see for example][]{Li+2015,SBM2015c,Sormani+2018b}. It is also worth nothing that the points where overshooting\footnote{We use the term ``overshooting'' to denote material that, after plunging towards the CMZ along one of the dust lanes, passes close to the CMZ but does not stop and continues towards the dust lane on the opposite side. See for example Figure 4 in \cite{Sormani+2019}.} material crashes into the dust lanes, which in \cite{Sormani+2019} we have interpreted as producing the observed extended velocity features (EVF), are sites of enhanced star formation. However, by the time this star formation is visible, these regions will have moved at high speed ($\sim 200\kms$) inwards towards the CMZ. The time delay between sink formation in our model and the star formation actually becoming visible will depend on our choice of star formation rate tracer, but we would expect it to be at least $\sim 0.4$~Myr (the free-fall time of the gas at the sink creation density). Star formation should become visible soon after this if observed with tracers that are insensitive to the dust extinction (e.g.\ radio recombination lines), or after a longer but poorly quantified period if observed with tracers such as H$\alpha$ that are highly sensitive to dust obscuration. This is consistent with observations of Bania Clump 2 (one of the most prominent EVF), which despite containing dozen of 1.1 mm clumps, has been found to be deficient in near- and mid-infrared emission in the Spitzer images and has been suggested to be in a pre-stellar stage of cloud evolution by \cite{Bally+2010}. Our simulations therefore support the idea that Bania Clump 2 will shortly begin to form massive stars.
 
A noteworthy feature of the averaged as well as of the instantaneous SFR density distributions (bottom panel of Figure~\ref{fig:average} and Figure~\ref{fig:SigmaSFR_v_time_CMZ}) is that there is a site of star formation inside the CMZ ring radius, after a radial gap. Indeed, we noted in Sections~3.4 and 5.2 of Paper I that gas can be found inside the CMZ radius in these simulations (in contrast to our previous non self-gravitating simulations in \citealt{Sormani+2019}, in which there was no gas inside the CMZ ring). This star formation might be associated with star formation occurring near SgrA* ($R \leq 10\pc$). This would be consistent with claims of observational evidence for ongoing star formation in this region \citep{YusefZadeh+2008,YusefZadeh+2015}, although we note that these claims are controversial at the moment \citep{Mills+2017}. Such star formation might also be related to the formation of the nuclear stellar cluster (NSC, see for example \citealt{Genzel2010,Schoedel+2014,GallegoCano+2020}) by providing in-situ newly born stars and, since such stars are rotating, it might contribute to its observed rotation \citep{Feldmeier+2014,Feldmeier+2015,Chatzopoulos+2015,Tsatsi+2017,Neumayer+2020}.

\begin{figure}
	\includegraphics[width=\columnwidth]{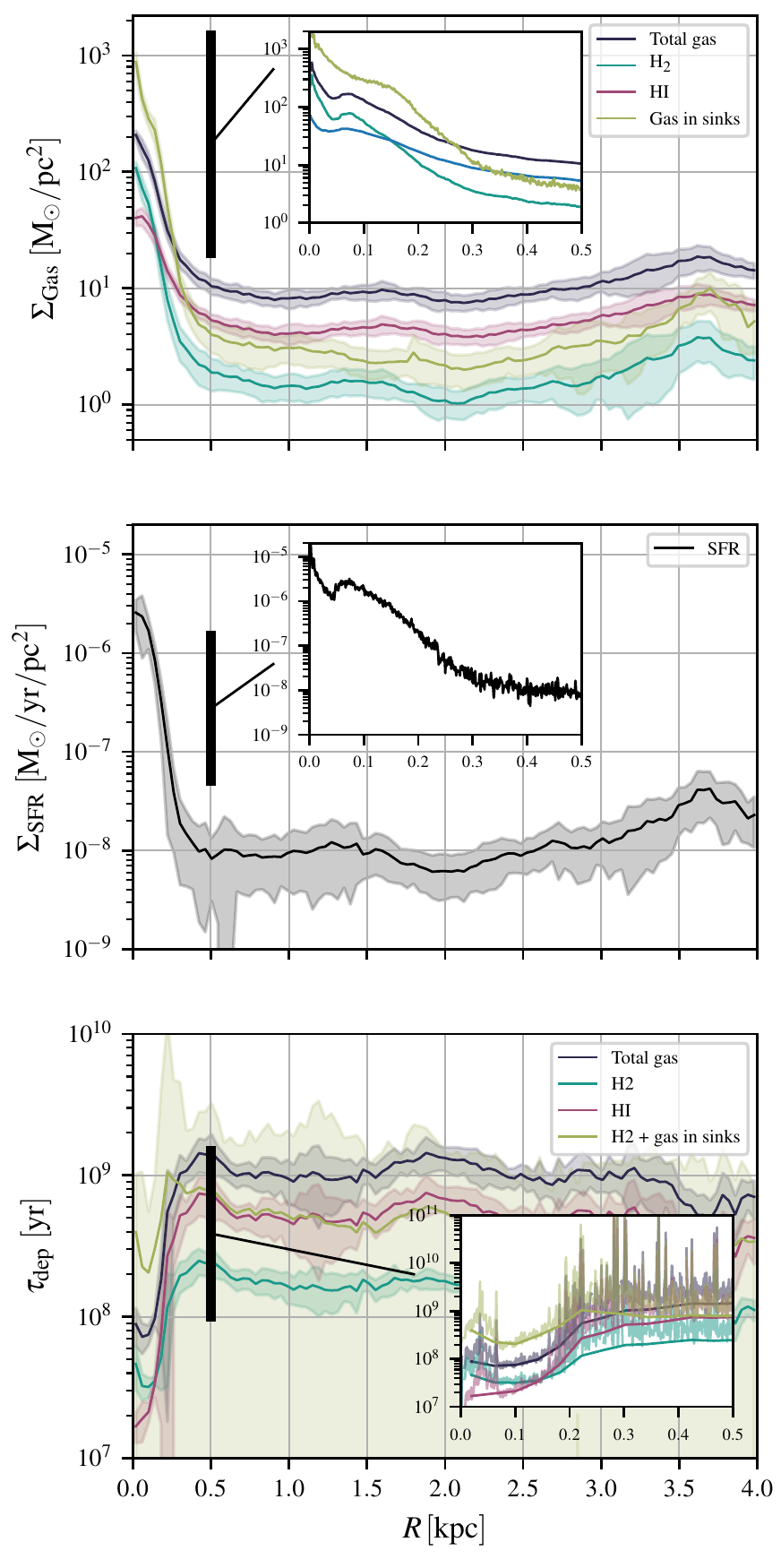}
    \caption{\emph{Top}: time-averaged radial distribution of gas surface density. \emph{Middle}: SFR density. \emph{Bottom}: depletion times. Plots are averaged over time in the range $t = (146.7,175.8) \Myr$. Shaded areas show the 1-sigma scatter. The zoom-in inlays show the time-averaged quantities in the innermost $0.5\kpc$ with a finer radial binning.}
    \label{fig:radialprofile}
\end{figure}

Figure \ref{fig:radialprofile} analyses the radial distribution of $\Sigma_{\rm gas}$, $\Sigma_{\rm SFR}$ and of the depletion time. The lines show the time-averaged values, while the shaded regions show the scatter. This figure indicates that both $\Sigma_{\rm gas}$ and $\Sigma_{\rm SFR}$ increase considerably in the centre, while the ratio between the two, the depletion time, decreases by a factor of $\sim 5$, consistent with what we found in Section~\ref{sec:temporalsf}. Indeed, the minimum of the depletion time is reached in the CMZ ring. 

Interestingly, the maximum of the depletion time as a function of radius is instead reached just outside the CMZ ring, at $R\simeq 500\pc$, in the terminal part of the dust lanes. This is where gas reaches the highest bulk speeds (and observed line of sight velocities) over the entire MW disc, and may indicate that star formation is suppressed at these sites due to the very high shear, in line with the arguments presented in \cite{Renaud+2015} and \cite{Emsellem+2015}. In order to check this, we plot in Figure \ref{fig:shear} the quantity 
\begin{equation} \label{eq:shear}
\tau=\left[ \left( \frac{\pa V_x}{\pa y} + \frac{\pa V_y}{\pa x} \right)^2 +  \left( \frac{\pa V_x}{\pa x} - \frac{ \pa V_y }{\pa y}\right)^2 \right]^{1/2}\, ,
\end{equation}
where $V_i = \int_\infty^\infty \rho v_i \di z / \int_\infty^\infty \rho \di z$ is the density-weighted projected velocity. The quantity $\tau$ is a good indication of shear for a 2D flow, and has the desirable property of being invariant under rotations of the coordinates since it is the magnitude of the eigenvalues of the traceless shear tensor $D_{ij}=\left[\pa_j V_i + \pa_i V_j - \delta_{ij} \left( \nabla \cdot \mathbf{V} \right)\right]/2$  \citep[e.g.][]{Maciejewski2008}. We estimate the derivatives $\pa_i V_j$ using finite differences with a resolution $\Delta x = 4\pc$, so any gradient on scales smaller than this is unresolved in the figure. Figure \ref{fig:shear} shows that indeed terminal parts of the dust lanes are regions of particularly high density and high shear (see red arrow in the figure), confirming our interpretation. 

\begin{figure*}
	\includegraphics[width=\textwidth]{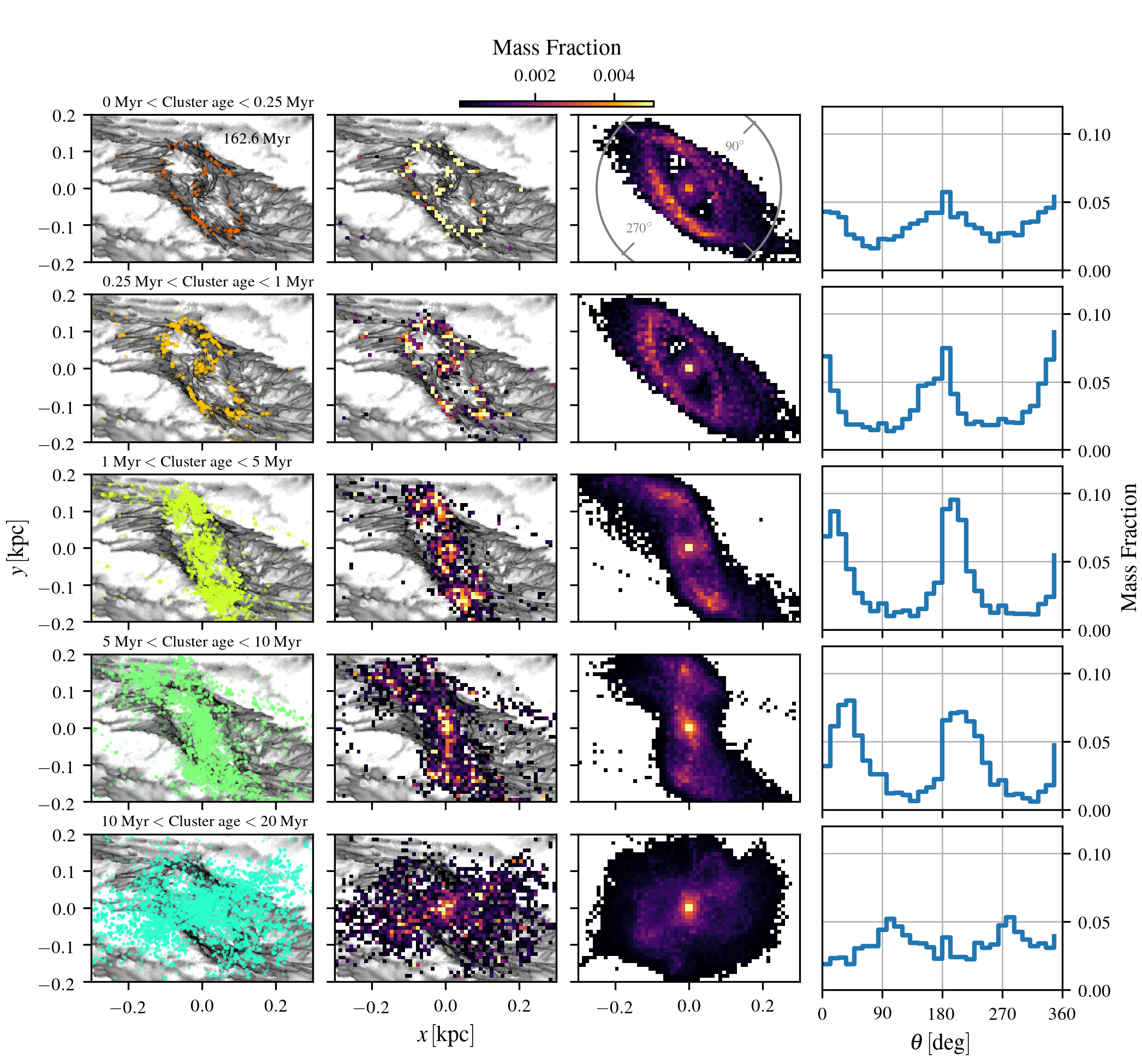}
    \caption{\emph{Left column}: instantaneous spatial distribution of stars with age in the given range, for a typical snapshot of our simulation. This is a scatter plot. \emph{Middle-left column}: same as the left panel, but binned in a 2D histogram weighted by mass. \emph{Middle-right column}: time-averaged spatial distribution of stars by age, i.e. obtained by time-averaging the middle-left column. The time averaged is taken over $t = (160, 180)$. \emph{Right column}: azimuthal distribution of stars with age in the given range in the CMZ ($R \leq 250\pc$), obtained by looking at the azimuthal distribution of the histograms in the middle-right column. This shows that the stars are not distributed uniformly through azimuth, but have distinct peaks whose azimuthal position depends on the age range of the stars considered.}
    \label{fig:starsbyage}
\end{figure*}

\begin{figure}
	\includegraphics[width=\columnwidth]{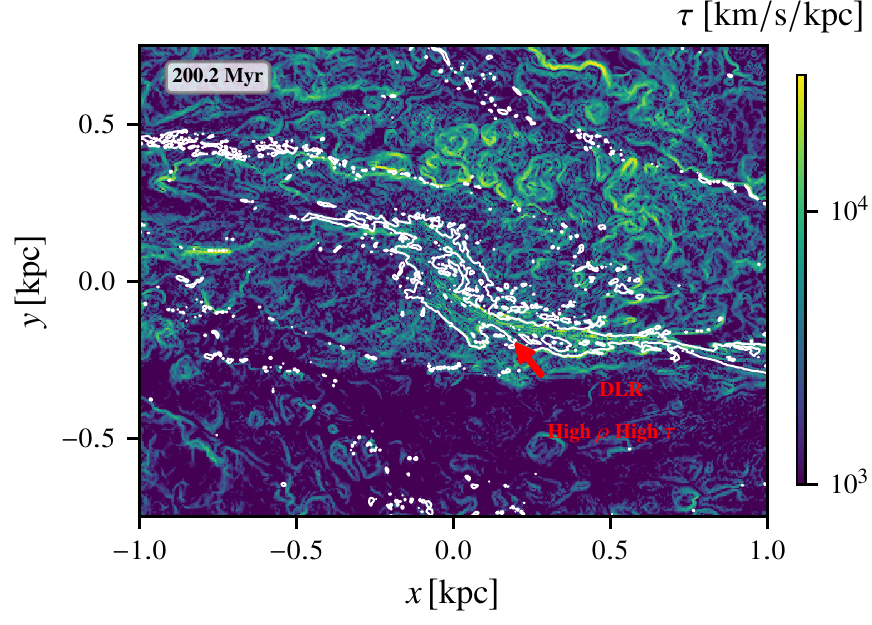}
    \caption{Shear map for the snapshot at $t=200.2\Myr$. The colours show the quantity $\tau$ defined in Equation \eqref{eq:shear}, which is a good indication of shear for a 2D flow. The white contour indicates where the total surface density of the gas is $N = 5 \times 10^{21} {\rm \, cm^{-2}}$. The red arrow indicates the terminal part of the dust lane, a region of high density and high shear, where the maximum of the depletion time as a function of Galactocentric radius is reached (see bottom panel in Figure \ref{fig:radialprofile} and discussion in Section \ref{sec:spatialsf}). The map also shows that high shear occurs predominantly in the dust lanes and in expanding SN shells.}
    \label{fig:shear}
\end{figure}

A more detailed analysis of the spatial distribution of star formation can be performed by subdividing the newly born stars into different age ranges. The left column in Figure~\ref{fig:starsbyage} performs this decomposition for an instantaneous snapshot, while the middle-right column shows the time-averaged version. One can see that the very young stars are well correlated with the dense gas, but they become increasingly decoupled as they age. Gas and stars have achieved significantly different spatial distributions by the time stars are $\sim$5 Myr old. The physical reason for the decoupling is as follows. Imagine a star and a gas element that are initially on the same orbit. In the CMZ, gas frequently collides with other gas (typically every $1\mhyphen 2\Myr$ and at least twice per orbit, when the CMZ gas collides with the dust-lane infall, see Paper I). In such a collision, the gas trajectory of the gas parcel will be strongly affected, while the star will simply fly through relatively undisturbed since it does not feel pressure forces according to its equations of motion. Therefore, after a few collisions the gas and the star will be on quite different trajectories. \cite{Renaud+2013} also noted decoupling between the stellar and gaseous component within spiral arms in their simulation (see their Section 4.5). However, in their case the decoupling was caused by asymmetric drift, i.e. by a lag between stars and gas caused by the larger velocity dispersion of stars compared to the gas, which plays a minor role in our case since it is overshadowed by the frequent collisions in the CMZ (which were absent in the dynamically quieter region studied by \citealt{Renaud+2013}).

\begin{figure}
	\includegraphics[width=\columnwidth]{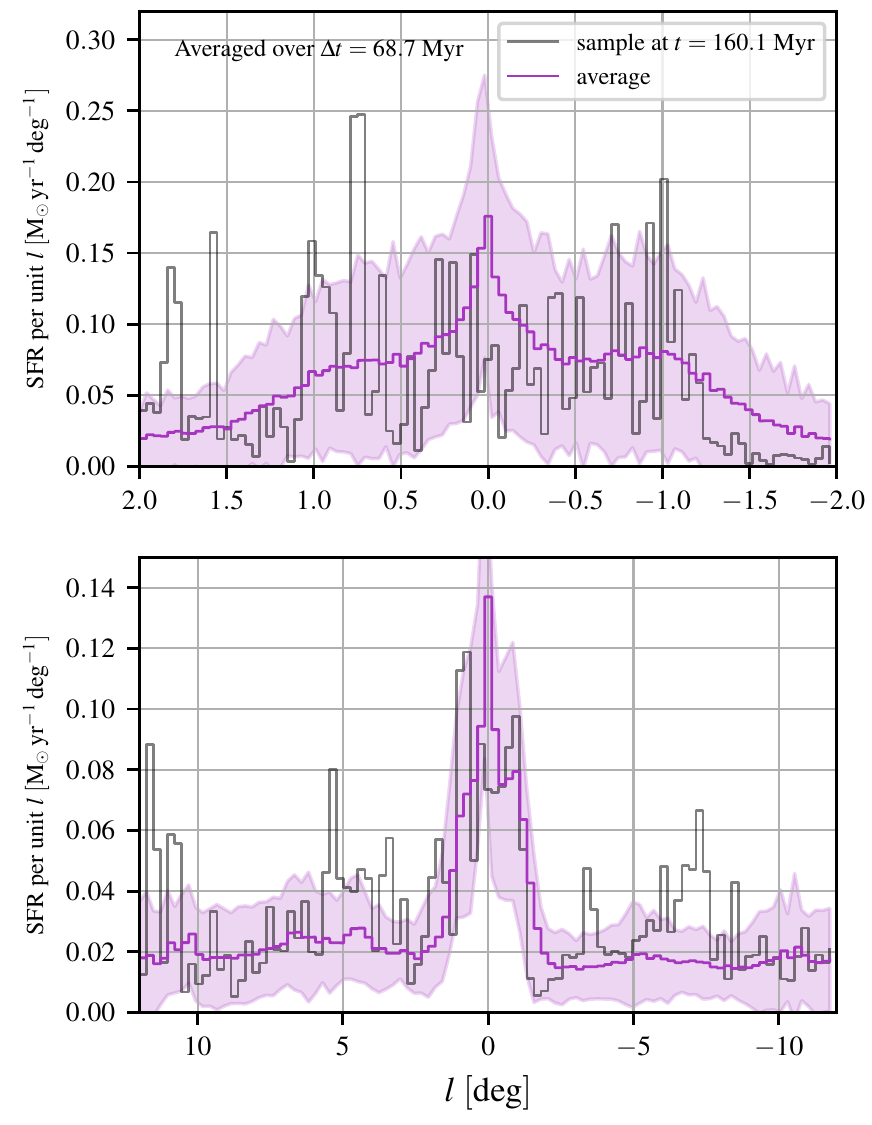}
    \caption{Star formation rate as a function of Galactic longitude in our simulation. The magenta line shows the time-averaged distribution, while the grey line shows the instantaneous distribution at $t=160.1 \Myr$. Shaded area shows the typical scatter. The time averages are calculated over the time range $t = (146.7, 215.4)$~Myr.} 
    \label{fig:sfr_long}
\end{figure}

Finally, we plot in Figure~\ref{fig:sfr_long} the SFR as a function of longitude. The averaged distribution has a large central peak and two smaller lateral peaks on the sides at $l \simeq 0.75^\circ$ and $l=-1^\circ$ (lower panel), roughly consistent with observations which have peaks at the position of SgrB2 and SgrC (see for example Figure~A1 of \citealt{Barnes+2017}). Again, fluctuations of the instantaneous distribution around the averaged distribution can be quite large, and the peaks can be more or less evident in the instantaneous distributions depending on the particular snapshot chosen.

\subsection{Trajectories of newly born stars} \label{sec:trajectories}

Once a sink particle is formed, it typically follows a different trajectory than the gas. As already noted in Section~\ref{sec:spatialsf}, this can be seen for example from Figure~\ref{fig:starsbyage}, which show how gas and stars in the CMZ quickly decouple and have achieved significantly different distribution within $\sim$5~Myr. As mentioned in that section, the main physical reason why stars and gas decouple is because gas trajectories are frequently disturbed by collisions, while stars continue on their path almost undisturbed.

\begin{figure}
	\includegraphics[width=\columnwidth]{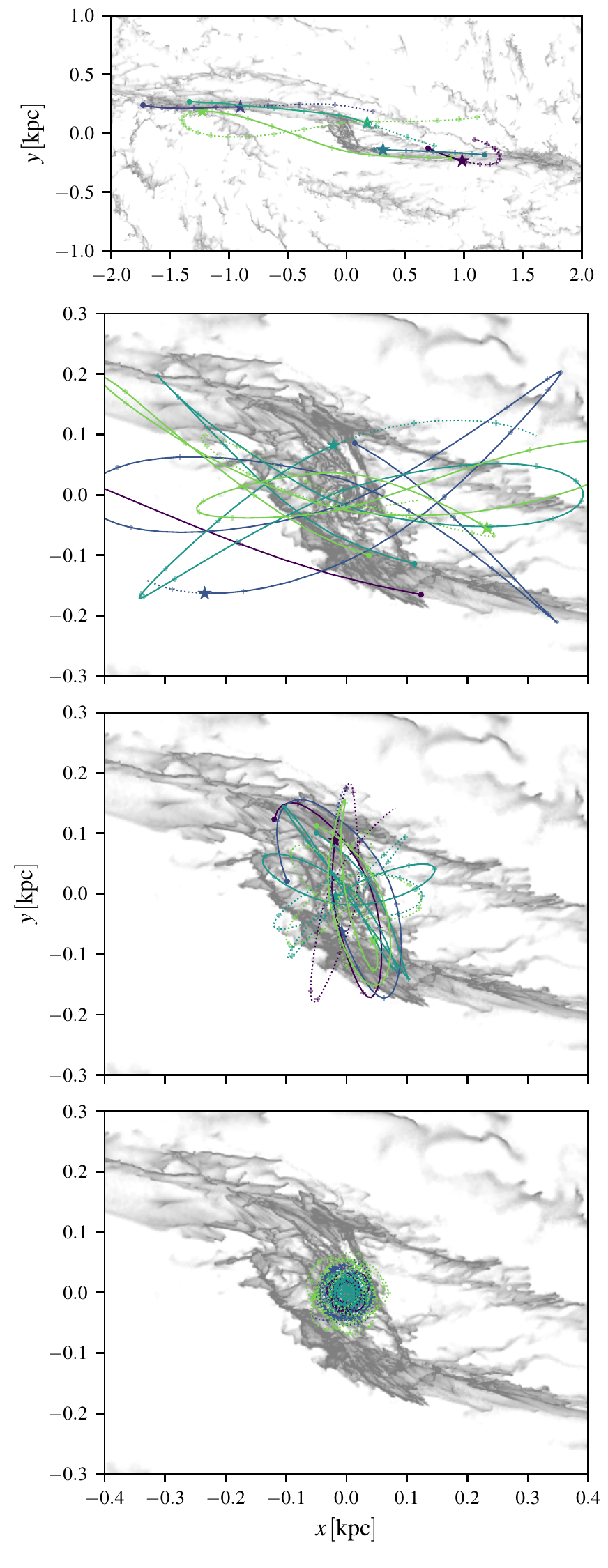}
    \caption{Typical trajectories of newly born stars in our simulations. \emph{Top panel}: for stars formed upstream along the dust lane. \emph{Middle-top panel}: for stars formed downstream along the dust lanes. \emph{Middle-bottom panel}: for stars formed from gas orbiting in the CMZ ring. \emph{Bottom panel}: for stars formed at radii smaller than the CMZ ring. Star markers indicate the current position of the stars. Round markers indicate the location where they formed. Full lines indicate the past trajectories, while dashed lines indicate the future trajectories. Cross markers log the position of the stars at equal time intervals of $\Delta t = 1 \Myr$. Grey shows the H$_2$ surface density at current time.}
    \label{fig:trajectories}
\end{figure}

Figure \ref{fig:trajectories} investigates the trajectories of a sample of sink particles in more detail. The first panel shows stars that are born upstream along the dust lanes, while the gas is on its way towards the CMZ. These stars will have very elongated orbits that often pass close to the centre with very high speed (up to $300 \kms$), and after each passage reach several kpc out from the centre. The second panel shows stars that formed downstream along the dust lanes, where the gas is accreting onto the CMZ. These stars will overshoot a little bit and typically have elongated orbits which are a factor of 2-3 larger than the CMZ ring. Typical orbital speeds of these stars are larger ($\sim 150 \kms$) than gas in the CMZ ring ($\sim 100 \mhyphen 120\kms$). The third panel shows stars formed within the CMZ ring. These stars will stay within the ring and have typical orbital velocities comparable to the gas in the ring ($\sim 100 \mhyphen 120\kms$), but after a few Myr they will decouple from the gas. The accumulation of stars similar to those shown in the second and third panel is what forms the nuclear stellar disc over time (NSD, see for example \citealt{Launhardt+2002,Nishiyama+2013,Schoenrich+2015,BabaKawata2020}). Finally, the last panel shows stars that have formed from gas inside the CMZ ring. These typically follow roughly circular orbits with moderate speeds ($\sim 80 \kms$), so they will remain inside the CMZ ring. As noted in Section \ref{sec:spatialsf}, such star formation might also be related to the formation of the nuclear stellar cluster (NSC, see for example \citealt{Genzel2010,Schoedel+2014,GallegoCano+2020}).

The trajectories of the sink particles in our simulation can be compared with the kinematics of star clusters and H{\sc II} regions. In Section~\ref{sec:arches} we compare them with the Arches and Quintuplet clusters. In an upcoming paper (Anderson et al., in preparation) we will compare them with H{\sc II} regions in the SgrE complex.

Finally, it is worth mentioning a limitation of our simulations. In the code, the gravitational force is calculated using a softening length, which for the gas is adaptive and depends on the cell size with a lower limit set at $0.1 \pc$, while for the sinks is constant at a value of $1 \pc$ (see Section 2.2 in Paper I). The finite length of the gravitational softening will introduce biases in the binding of stellar structures. Thus, while we are able to retrieve the average motion of a small group of stars, we cannot properly retrieve the velocity distribution of individual stars.

\section{Discussion} \label{sec:discussion}

\subsection{What drives the time variability of the SFR in the CMZ?} \label{sec:sfrate}

The current global SFR in the CMZ (intended here as the region within $R \lesssim 200 \pc$, or $|l| \lesssim 1.4 \degree $ assuming a distance to the Galactic centre of $8.2\kpc$, e.g. \citealt{Reid+2019,Gravity2019}) is of the order of $\simeq 0.1 \Msunyr$ \citep[e.g.][]{YusefZadeh+2009,Immer+2012,Longmore+2013a,Barnes+2017}. This number is obtained by combining different independent methods, including direct counting of young stellar objects and integrated light measurements. All these methods agree with each other within a factor of two (see Table 3 in \citealt{Barnes+2017}), and also agree with the number obtained from counts of supernovae remnants (see Section 8.9 in \citealt{Ponti+2015}). Since these various methods trace SF over different timescales in the range $0.1 \mhyphen 5\Myr$, this also implies that the SFR in the CMZ has been roughly constant for the past $\sim 5 \Myr$ \citep{Barnes+2017}. Considering longer timescales, \cite{NoguerasLara+2020} has recently studied the star formation history in the CMZ region by modelling the extinction-corrected K-band color-magnitude diagram as a superposition of star formation events at different times. They found that the SFR averaged over the past $30 \Myr$ is $0.2 \mhyphen 0.8 \Msunyr$, i.e. a factor of a few higher than the rate averaged over the last $5 \Myr$. They also found that the SFR has been variable during the past Gyr, with periods of more intense activity ($\sim 0.5 \Msunyr$). This suggests that the SFR in the CMZ is not constant, but varies in time. Evidence for time variability in the star formation activity has also been found by \cite{Sarzi+2007} for external galactic nuclei by analysing the star formation history of a sample of nuclear rings. It is therefore natural to ask: what drives the time variability in the SFR of the CMZ?

A possible explanation is that the CMZ goes through episodic starbursts driven by feedback instabilities \citep{KrumholzKruijssen2015,Krumholz+2017,Torrey+2017,Armillotta+2019}. In this scenario, the CMZ has a roughly constant gas mass but order-of-magnitude level variations in the SFR. The depletion time is not constant, but has large variations over time. The large scatter ($ \sim 1$ dex) in the depletion times observed in the centre of external barred galaxies \citep{Leroy+2013,Utomo+2017} is explained by temporal fluctuations. \cite{Armillotta+2019} run numerical hydrodynamical simulations of gas flowing in a barred potential which included star formation prescriptions that lend support to this scenario. In their simulation, the CMZ depletion time is not constant, and SFR variations are driven by variations in the depletion time rather than by variations in the mass of the CMZ.

Our simulations suggest an alternative scenario. Contrary to the findings of \cite{Armillotta+2019}, we do not find that the CMZ depletion time goes through strong oscillatory cycles. Instead, our simulation predicts that the depletion time is approximately constant in time (within a factor of two, see Section~\ref{sec:temporalsf} and Figure~\ref{fig:tdepl}), so that the SFR is roughly proportional to the total mass of the CMZ. This suggests that variations in the SFR reflect changes in the mass of the CMZ rather than changes in the depletion time/SF efficiency. Fluctuations in the mass of the CMZ could come from a variety of factors that are not included in our simulation. For example, the mass of the CMZ might drastically and suddenly decrease due to gas expulsion caused by AGN feedback. Perhaps, an AGN event associated with the Fermi Bubbles \citep{Su+2010} is what caused the observed drop in the SFR from the value $0.2\mhyphen 0.8 \Msunyr$ $\sim 30\Myr$ ago \citep{NoguerasLara+2020} to the value $\sim 0.1 \Msunyr$ inferred for the last $5\Myr$ \citep{Barnes+2017}. This would be compatible with the currently estimated ages of the Fermi Bubbles (see for example \citealt{Mou+2018} and references therein). The mass of the CMZ could also change due to variations in the accretion rate, induced for example by an external perturbation such as a merger. We note that at the current estimated mass inflow rate of $\sim 1 \Msunyr$ \citep{SormaniBarnes2019}, the entire current gas mass of the CMZ ($\simeq 5 \times 10^7 \Msun$) can be accumulated in just $50 \Myr$, so a change in this rate could potentially induce mass and SFR variability within the timescales required by observations. We also note that much higher accretion rates seem to be possible in barred galaxies: for example \cite{Elmegreen+2009} reports a bar-driven inflow rate of $40 \Msunyr$ in NGC 1365. Our scenario is also supported by the work of \cite{Seo+2019}, who run hydrodynamical simulations of gas flowing in a live N-body barred potential and find that the SFR correlates well with the bar-inflow rate. In our scenario, the large scatter in the depletion times observed in the centre of external barred galaxies \citep{Leroy+2013,Utomo+2017} would be explained as due to different environmental conditions rather than to high time variability. For example, different strengths of the stellar gravitational potential might contribute to the scatter in the depletion times (see Equation \ref{eq:ostrikershetty} and related discussion). Note also that some of the scatter in these values may be driven by differences in the size of the CMZ-like region in different galaxies, since this region is typically not resolved in the kpc-scale molecular gas maps considered in \citet{Leroy+2013} and \citet{Utomo+2017}.

What causes the differences between the results presented here and those in \cite{Armillotta+2019}? There are several factors that could contribute to this and it is difficult to point to which one is most important. First, the two papers use significantly different treatments of ISM cooling. \citet{Armillotta+2019} treat gas cooling using equilibrium cooling curves provided by the {\sc grackle} astrochemistry and cooling package \citep{Smith+2017}, which potentially yield differences in behaviour compared to the fully non-equilibrium treatment we use here. In addition, they treat photoelectric heating as a uniform heating process and do not account for variations in the heating rate due to changes in the fractional ionisation of the gas or its degree of dust shielding. Although the two treatments result in ISM phase diagrams that are qualitatively similar in many aspects (compare their Figure~5 with Figure~11 in Paper I), there are clear quantitative differences that may have some impact on the predicted star formation rates. 

Second, the star formation prescription used in \citet{Armillotta+2019} is also quite different from that used in our code. In their approach star particles are stochastically formed in gas denser than $10^{3} \: {\rm cm^{-3}}$, provided that it is gravitationally bound, cold and self-shielded. Compared to our scheme, the main differences are their choice of density threshold and the fact that in their scheme, significant quantities of dense gas can accumulate above the density threshold, something which is impossible by design in our scheme. Third, \citet{Armillotta+2019} include the effects of photoionisation feedback as well as supernova feedback, while we concentrate here solely on the latter. 

Finally, there is a substantial difference in the mass resolution achieved in dense gas in the two simulations. In our simulation, gas at densities around $10^{3} \: {\rm cm^{-3}}$ is typically resolved with Voronoi cells with a mass of around $2\Msun$ (see Figure~3 in Paper I). In contrast, the default particle mass in \citet{Armillotta+2019} is $2000\Msun$, a factor of 1000 worse than we achieve here.  \citet{Armillotta+2019,Armillotta+2020} also present results from a ``high resolution'' run with a particle mass of $200\Msun$, which they carried out for a much shorter period than their main run, but even this has a much worse resolution than our simulation. An important consequence of this difference in resolution is that in the \citet{Armillotta+2019} simulation, the Sedov-Taylor phase following a supernova explosion is resolved only for supernovae exploding in low density gas with $n < 1 \: {\rm cm^{-3}}$, whereas in our simulation it remains well-resolved even for supernovae exploding in gas with a density close to our sink creation threshold. Therefore, \citet{Armillotta+2019} primarily inject momentum with their supernovae, since the associated thermal energy is rapidly radiated away, whereas we are able to follow the injection of both thermal energy and momentum in a more self-consistent fashion. This results in a clear difference in the morphology of the supernova-affected gas: in our simulation, supernova explosions produce large holes in the gas distribution, while corresponding features are rarely seen in the \cite{Armillotta+2019} simulation. 

In view of these significant differences in numerical approach, together with the fact that the results \citet{Armillotta+2019} obtain for the star formation rate of the CMZ are clearly not numerically converged (see their Figure A2), and that the simulations span a quite different period in the life of the CMZ ($\sim 100\Myr$ after bar formation in our run vs.\ $500 \Myr$ in their simulation), it is difficult to assess the reasons for the difference in results regarding the time variability of the SFR in the CMZ. This is an issue that we hope to address further in future work.

\subsection{An evolutionary sequence of star formation?} \label{sec:timelinesf}

\cite{Longmore+2013b} and \cite{Kruijssen+2015,Kruijssen+2019} suggested that star formation follows an evolutionary timeline as the gas clouds orbit the CMZ ring. In this scenario, star formation is triggered when the clouds are compressed during pericentre passage, i.e. when the clouds pass closest to the Galactic centre. This scenario is at variance with the two scenarios for star formation in nuclear rings that are more commonly discussed in the extragalactic context, namely the ``popcorn'' and the ``pearls on a string'' scenarios, which are schematically depicted in Figure~7 of of \cite{Boker+2008}. In the ``pearls on a string'' scenario, star formation occurs prevalently at the contact point between the dust lanes and the gas ring, which typically coincides with the ring apocentre rather than with the pericentre. In the ``popcorn'' scenario, star formation occurs uniformly along the ring. The observational evidence for a clear evolutionary sequence as implied by the pericentre passage scenario is mixed \citep{Kauffmann+2017b,Krieger+2017}, while the peals on a string scenario has obtained some mild support from observations of nearby galaxies (see for example Section~4.1 in \citealt{Boker+2008}, see also \citealt{Mazzuca+2008}).

These three scenarios make different predictions that can be tested with our simulations. The pericentre scenario predicts that star formation occurs predominantly after the pericentre passage. According to this scenario, very young stars should be found shortly after the passage, while stars of increasing age should be found further downstream of the pericentre. The pearls on a string scenario predicts that star formation happens predominantly downstream of the contact point between the dust lanes and the CMZ ring, i.e. downstream of the apocentre. The popcorn scenario predicts that star formation is distributed uniformly along the ring, without preferred locations.

In order to test these predictions, we look at the time-averaged distribution of very young stars (age $t \leq 0.25 \Myr$), which is shown in the top row of Figure~\ref{fig:starsbyage}. These stars trace where the star formation is being triggered. The right panel in the top row shows the azimuthal distribution of stars in the CMZ. The apocentres of the CMZ ring are at $\theta=0$ and $\theta=180^\circ$, and coincide with the contact points between dust lanes, while the pericentres are at $\theta=90^\circ$ and $\theta=270^\circ$. This panel shows that the distribution of very young stars has a bi-periodic structure with two strong peaks at the apocentres, consistent with the prediction of the pearls on a string scenario.

\begin{figure*}
	\includegraphics[width=\textwidth]{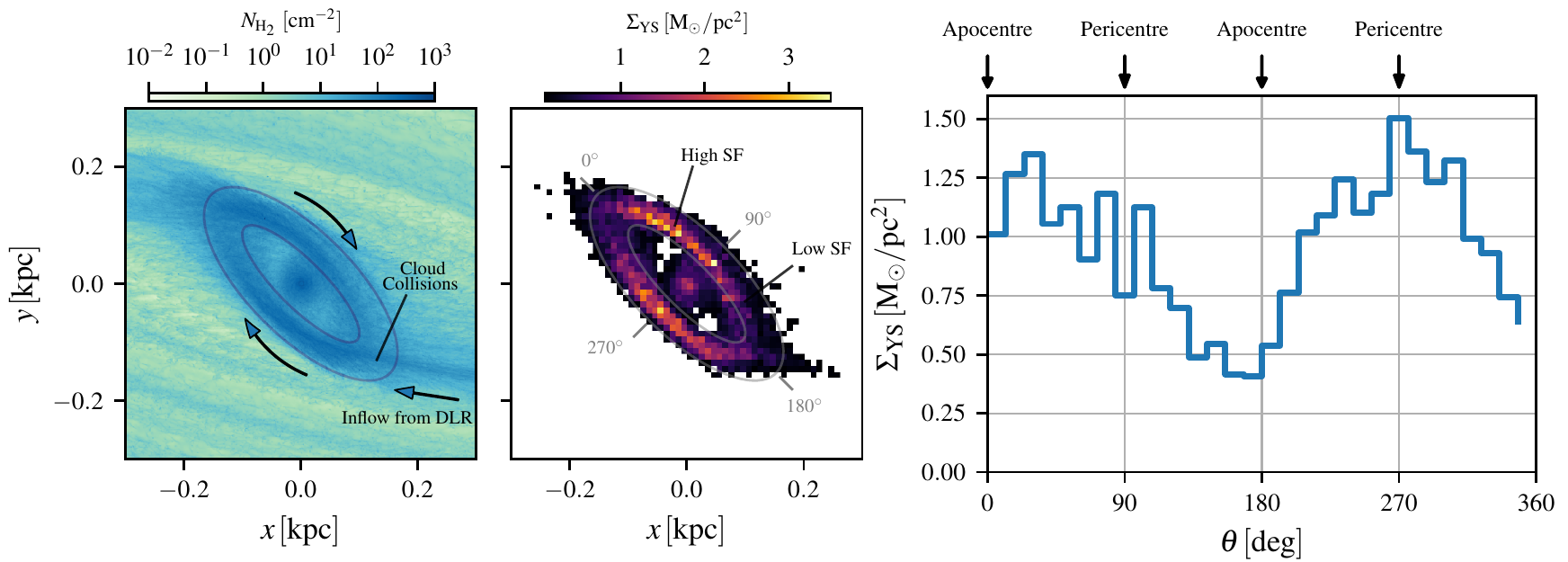}
    \caption{\emph{Left}: time-averaged H$_2$ surface density. \emph{Middle}: time-averaged surface density of the very young stars (age $t \leq 0.25 \Myr$) formed in our simulation. \emph{Right}: time-averaged surface density of very young stars (age $t \leq 0.25 \Myr$) as a function of azimuth along the elliptical ring shown in the middle panel. The time averages are calculated over the time range $t = (146.7, 168.7)$~Myr. }
    \label{fig:azimuthal}
\end{figure*}

The above analysis considers all the star formation within $R \leq 250 \pc$, including some that strictly speaking is outside the ``ring'' structure. In order to investigate this aspect in more detail, we focus specifically on the ring in Figure \ref{fig:azimuthal}. The right panel shows the time-averaged surface density\footnote{By plotting the \emph{surface density} rather than a histogram of the mass distribution as a function of azimuth, we avoid any potential bias due to geometric effects caused by the area within the ellipse not being constant in each angular range. For example, if the surface density were constant along the ring, the azimuthal distribution of mass would not be constant, although the 2D face-on maps would look perfectly uniform.} of very young stars ($\Sigma_{\rm YS}$) within the elliptical ring shown in the middle panel. It can be seen that most of the star formation occurs downstream of the apocentres, but before the pericentre passage. This is consistent with the prediction of the pearls on a string scenario, but not with the pericentre passage scenario. Note however that the maxima are quite broad, and star formation away from these maxima is not zero. Thus, while the maxima constitute a region of more intense star formation, they are not the only regions where stellar birth takes place.

Let us now consider the distribution of older stars, which can be seen from Figure~\ref{fig:starsbyage}. For ages $1 < t < 5 \Myr$, the distribution of stars in the middle-right column exhibits a clear bi-polar structure. This is because stars accumulate close to the apocentre, where their orbital velocity slows down and where they therefore spend more time than in other parts of their orbit. As stars become older ($t > 10 \Myr$), the bipolar structure precesses as a consequence of the precession of the apocentres of the stellar orbit (which at their formation prevalently coincide with the contact point between ring and the dust lanes, but change at later times). The bipolar structure also becomes less pronounced, and the distribution more uniform, as stars mix in phase space

We remark that all our conclusions above come from analysing the \emph{time-averaged} distributions. As discussed in Section~\ref{sec:spatialsf}, the instantaneous star formation distribution fluctuates strongly around the average (compare the left panels in Figure~\ref{fig:starsbyage} and the various panels in Figure~\ref{fig:SigmaSFR_v_time_CMZ} with the middle-right panels of Figure~\ref{fig:starsbyage}). Because of these fluctuations, it is much harder to tell whether our simulations are consistent with the pearls on a string scenario by looking just at a single snapshot. Moreover, while the time-averaged distributions favour the pearls on a string scenario, there is also significant star formation throughout the ring and away from the apocentres. These complications should be taken into account when analysing observations, which only constitute individual snapshots. 

From a physical point of view, there are two reasons why enhanced star formation should be expected at the apocentres: (i) they are collisions sites where the gas from the dust lanes crashes into the ring (see e.g.\ the left panel in Figure~\ref{fig:azimuthal}); (ii) gas slows down at the apocentre of an orbit, causing it to pile up and become more dense. Our simulations suggest that these effects are dominant over the tidal compression at the pericentre proposed by \cite{Kruijssen+2015}. Even neglecting these two dominant effects, there is evidence that the pericentre passage only has a minor role in triggering star formation events. We note that in the simulations of \cite{Dale+2019} the pericentre has a rather weak effect in enhancing the SFR (compare the circular and non-circular orbits in Figures 3 and 9 of \citealt{Dale+2019}). \cite{Jeffreson+2018} also estimates that only a small fraction ($\sim 20\%$) of the star formation events might be triggered by pericentre passage. Their estimate neglects the two dominant mechanisms mentioned above, i.e. cloud collisions at the dust lanes and gas slowing down at the apocentre, so it is likely that the actual number is significantly lower than this. Finally, \cite{Kruijssen+2019} also acknowledge that star formation might be triggered by accretion, similarly to the pearls-on-a-string scenario. However, in their discussion the accumulation of gas in the CMZ takes place within the context of the \cite{KrumholzKruijssen2015} model rather than from direct accretion from the dust lanes. As we have argued in Section 6.2 of Paper I, the theoretical framework of \cite{KrumholzKruijssen2015} and \cite{Krumholz+2017} does not capture well the physics of the CMZ since it predicts the existence of a quasi-axisymmetric outer CMZ extending out to $R\simeq450\pc$, which is not supported by either observations or simulations. Moreover, \cite{SormaniLi2020} have shown that the acoustic instability on which these models are based is a spurious result which cannot drive turbulence and mass transport in the interstellar medium.

We conclude that our simulation supports a scenario which is a mixture of the pearls on a string and of the popcorn scenarios. Most of the star formation happens downstream of the apocentres, but a significant amount of star formation also takes place distributed along the ring. Our results do not support the pericentre passage scenario.

\subsection{Star formation relations} \label{sec:sfrelations}

Star formation relations are empirical correlations between the SFR and properties of the interstellar medium (ISM) from which stars are born. It has been extensively discussed in the literature that the CMZ follows some star formation relations but not all of them \cite[e.g.][]{YusefZadeh+2009,Longmore+2013a,Kruijssen+14b,Kauffmann+2017a,Kauffmann+2017b}. In particular, it has been shown that the \emph{global} SFR of the CMZ is consistent with the Schmidt-Kennicutt density relation \citep{Schmidt1959,Kennicutt1998}, with the \cite{Bigiel+2008} molecular gas relation, and with the \cite{Bacchini+2019a,Bacchini+2019b} volumetric star formation relation (Bacchini, private communication). However, the global SFR of the CMZ is not consistent with the SFR-dense gas relation observed by e.g.\ \citet{GaoSolomon2004}, \citet{Wu+2005} and \citet{Lada+2010,Lada+2012}. This is a linear relation between the quantity of dense gas (as traced by HCN emission or high dust extinction) and the SFR. It has been shown to work well both for the total (integrated) properties of external galaxies, and for local molecular clouds in the MW, which made it apparently valid over an impressive 9 orders of magnitude (although with a gap in the middle, see Figure~2 in \citealt{Lada+2012}). This generated the expectation that the same law should be valid for the CMZ, but the data shows that it is not (see \citealt{Longmore+2013a,Kruijssen+14b} and Figure~1 in \citealt{Kauffmann+2017b}). This expectation, and the universality of the SFR-dense gas relation, is also challenged by observations that suggest that the centres of nearby galaxies lie on average below the \citealt{Lada+2012} relation (see \citealt{Gallagher+2018,JimenezDonaire+2019} and in particular Figure~13 in the latter).

\begin{figure}
	\includegraphics[width=\columnwidth]{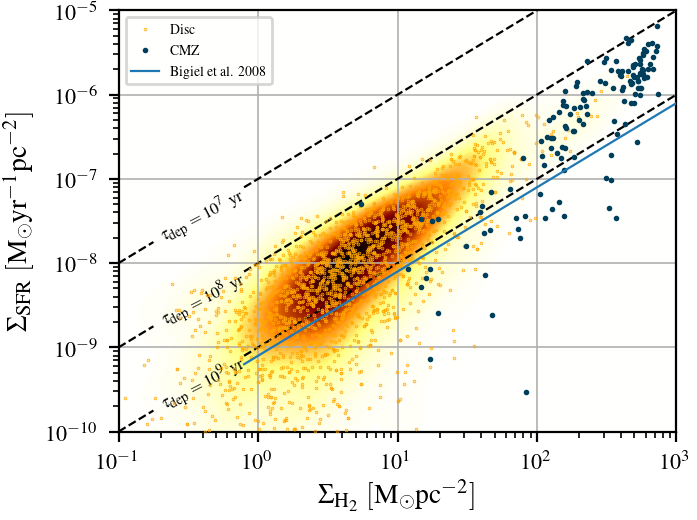}
    \caption{Schmidt-Kennicutt plot for our simulation. We bin face-on H$_2$ and SFR surface densities with a grid size of $100$~pc. Each point in this graph represents one such bin. The points are coloured based on the position of the centre of the bin. The underlying distribution is obtained by Gaussian kernel density estimation of the points associated to the disc. To increase statistics especially for the CMZ we include surface densities of eight different consecutive snapshots i.e. over a time of approximately $2$~Myr. The CMZ approximately follows the SK as well as the \citet{Bigiel+2008} relation, as found in observations \citep[e.g. Figure~2 in][]{Kruijssen+14b}.}
    \label{fig:SK}
\end{figure}

Figure \ref{fig:SK} shows that the CMZ in our simulation follows the Schmidt-Kennicutt relation \citep{Schmidt1959,Kennicutt1998} and the \cite{Bigiel+2008} molecular gas relation, consistent with observational findings. This reassures us that our numerical star formation subgrid model is working correctly. Unfortunately, our simulations do not have the resolution to probe the dense-gas star formation relations, which the CMZ has been shown to be not consistent with \cite[e.g.][]{Longmore+2013a,Kruijssen+14b,Kauffmann+2017a,Kauffmann+2017b}. To do that, we would need to increase the sink formation density threshold $\rho_{\rm c}$ (see Appendix \ref{sec:resstudy}) to densities of $n \simeq 10^7 \cmthree$, which is the dense gas formation threshold in the CMZ estimated by \cite{Kruijssen+14b} and \cite{Kauffmann+2017b}. This 
is impractical with our current simulations owing to the very high computational expense, but is a worthwhile direction for future investigations.

\subsection{The Arches and Quintuplet clusters} \label{sec:arches}

\begin{figure}
	\includegraphics[width=\columnwidth]{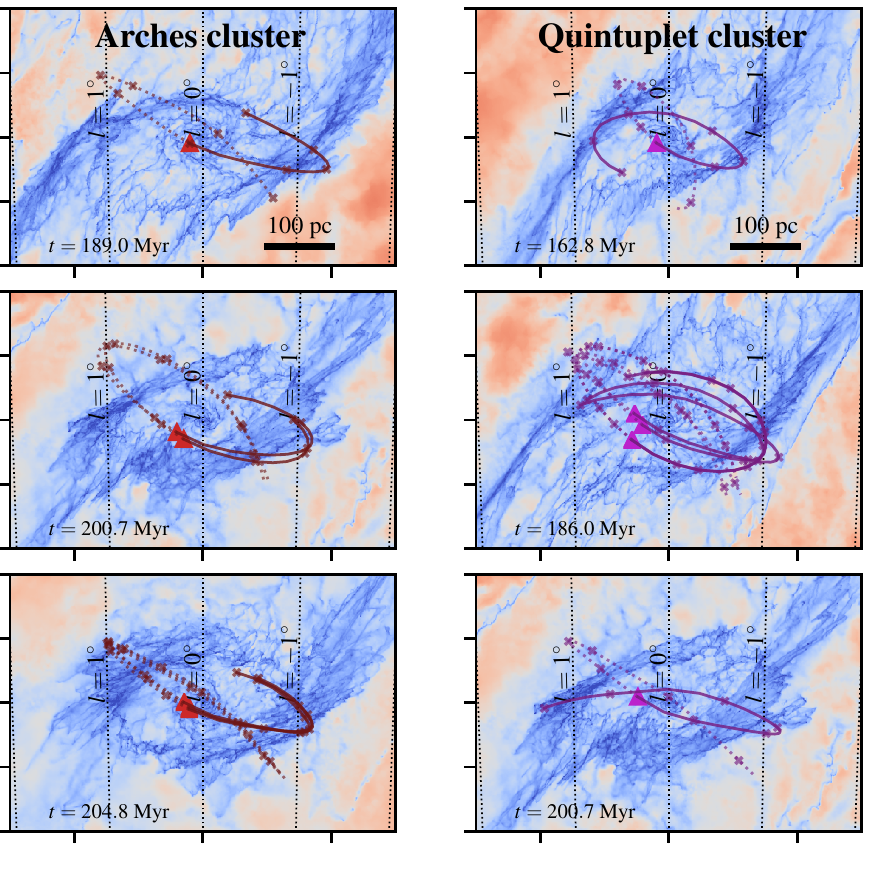}
    \caption{Sink particles in our simulations with properties (age, line-of-sight velocity, proper motion velocity) within the observational constraints of the Arches {\em (left panels)} and the Quintuplet \emph{(right panels)} clusters. Red/violet triangles denote the present day position, while solid and dotted lines show the past (from the birth site to the current position) and the future trajectories (for the next 5 Myr) respectively. The crosses log the position of the cluster at equal time intervals of $1$~Myr. The background shows the gas total density distribution at the time when the clusters are at their present day position.}
    \label{fig:arches_quintuplet}
\end{figure}

The Arches and Quintuplet clusters are two young massive ($M \gtrsim 10^4 \Msun$) clusters found close to the Galactic centre ($\simeq 30 \pc$ in projected distance). They have estimated ages of $3.5 \pm 0.7 \Myr$ and $4.8 \pm 1.1 \Myr$ respectively \citep{Schneider+2014}. The Arches cluster has a line-of-sight velocity of $v_{\rm los} = 95 \pm 8 \kms$ \citep{Figer+2002} and a proper motion velocity of $v_{\rm pm} = 172 \pm 15 \kms$ \citep{Clarkson+2012}, which yields a 3D orbital velocity of $v_{\rm 3D} = 196 \pm 17 \kms$ in the direction of increasing longitude \citep{Clarkson+2012}. The Quintuplet cluster has a line-of-sight velocity of $v_{\rm los} = 102 \pm 2 \kms$ and a proper motion velocity of $v_{\rm pm} = 132 \pm 15 \kms$, which yields a 3D orbital velocity of $v_{\rm 3D} = 167 \pm 15 \kms$, also in the direction of increasing longitude \citep{Stolte+2014}.

The observed motions of the Arches and Quintuplet clusters can be compared with the trajectories of our sink particles discussed in Section~\ref{sec:trajectories}. We have searched in our simulations for sink particles that are within $30\pc$ of the Galactic centre (in projected distance) on the positive longitude side and that have age, line-of-sight velocity and proper motion velocities compatible with those of the observed clusters within the observational uncertainties given above. Figure~\ref{fig:arches_quintuplet} shows trajectories for a sample of sinks which are found according to this procedure. This figure suggests that (i) the Arches cluster (left panels) formed from gas that is colliding into the far side dust lane at negative longitudes while orbiting in the CMZ. All the clusters in our simulation compatible with the above observational constraints are consistent with this picture. (ii) The Quintuplet cluster formed either in a similar scenario as the Arches cluster, but from gas colliding onto the near side dust lane (top-right panel), or more probably by gas in the terminal part of the dust lanes which is just entering the CMZ (middle- and lower-right panels). Occurrences of the second type are more frequent (roughly by a factor of $~\sim 5$).

Comparing with other works, the scenario described here is in some respects similar to the one proposed by \cite{Stolte+2008,Stolte+2014}, according to which the clusters are formed on a transitional trajectory between $x_1$ and $x_2$ orbits, since this transition happens at the contact point between the dust lanes (compare Figure~\ref{fig:arches_quintuplet} with Figure~12 in \citealt{Stolte+2014}). \cite{Kruijssen+2015} have proposed that the clusters originated on the same orbit that they use to fit dense gas data. However, we find that sink particles with properties compatible with the observed kinematics of the clusters have typically decoupled from the gas in which they are born by the time the clusters have reached their present age. Moreover, the gas orbiting in the CMZ ring has typically lower absolute 3D velocities than those of the clusters. Therefore, the scenario proposed by \cite{Kruijssen+2015} seems to be inconsistent with the result of the present simulation.

\section{Summary and conclusions} \label{sec:conclusion}

We have used the high-resolution hydrodynamical simulations presented in Paper I to study star formation in the CMZ. These include a realistic Milky Way external barred potential, a time-dependent chemical network that keeps track of hydrogen and carbon chemistry, a physically motivated model for the formation of new stars using sink particles, and supernovae feedback. The simulations reach sub-parsec resolution in the dense regions and allow us to resolve individual molecular clouds which are formed self-consistently from the large-scale flow.

Our main conclusions are as follows:
\begin{itemize}
\item We have studied the temporal distribution of star formation. We find that the depletion time in the CMZ is approximately constant in time. This implies that variations in the SFR of the CMZ are primarily driven by variations in its mass, caused for example by changes in the bar-driven inflow rate, AGN events or other external factors, while the observed scatter in the depletion time of external galactic centres is interpreted as variations in the environmental factors (e.g. the stellar surface density, \citealt{JimenezDonaire+2019}). Contrary to the findings of \cite{Armillotta+2019}, we do not find that the depletion time in the CMZ goes through strong oscillatory cycles, at least within the timescale of our simulation ($\sim 100 \, {\rm Myr}$, see Sections~\ref{sec:temporalsf} and \ref{sec:sfrate}).
\item We have studied the spatial distribution of star formation. Most of the star formation happens in the CMZ ring at $R\gtrsim100 \pc$, but a significant amount of star formation also occurs closer to SgrA*  ($R \leq 10\pc$, see Section \ref{sec:spatialsf} and Figure \ref{fig:SigmaSFR_v_time_CMZ}). While the time-averaged spatial distribution of the SFR is typically smooth, the instantaneous distribution can have complex and transient fluctuations which deviate significantly from the average morphology (compare the bottom panel in Figure \ref{fig:average} with Figures \ref{fig:SigmaSFR_v_time} and \ref{fig:SigmaSFR_v_time_CMZ}). Molecular clouds formed self-consistently from the large-scale flow, and their embedded star formation, exhibit complicated filamentary morphologies and do not resemble the idealised ``spherical clouds'' that are often used as a model to understand star formation.  We have also investigated how the spatial distribution changes when we consider stars in different age ranges, and found that a bi-polar structure persists even for stars with age $10\mhyphen20 \Myr$ (see Section~\ref{sec:spatialsf} and Figure~\ref{fig:starsbyage}).
\item We tested the predictions of the three main scenarios that have been put forward to explain the spatial and temporal distribution of star formation in the centre of barred galaxies, namely the ``pearls on a string", the ``popcorn" and the ``pericentre passage" scenarios. We found that our simulations are consistent with a mixture of the pearls on a string and popcorn scenarios, while they are inconsistent with the pericentre passage scenario (see Section~\ref{sec:timelinesf}).
\item We have studied the trajectories of newly born stars (see Figure \ref{fig:trajectories}). We find that gas and stars typically decouple within at most $2\mhyphen 3 \Myr$ (see Sections \ref{sec:spatialsf} and \ref{sec:trajectories}).
\item We have used the trajectories of newly born stars to provide a detailed analysis of the origin of the Arches and Quintuplet clusters. Our simulation favour a scenario in which the Arches cluster is formed from gas that crashed into the far side dust lane at negative longitudes while orbiting in the CMZ, while the Quintuplet cluster is either formed in a similar event but with the roles of the near/far sides the Galaxy reversed, or more likely by gas in the terminal part of the near side dust lane which was just entering the CMZ (see Figure~\ref{fig:arches_quintuplet} and Section~\ref{sec:arches}).
\end{itemize}

\section*{Acknowledgements}

We thank the anonymous referee for constructive comments that improved the paper. MCS thanks Ashley Barnes, Adam Ginsburg, Jonathan Henshaw, Zhi Li, Diederik Kruijssen, Alessandra Mastrobuono-Battisti, Elisabeth Mills, Nadine Neumayer, Francisco Nogueras-Lara, Ralph Schoenrich and Alessandro Trani for useful comments and discussions. MCS and RGT both acknowledge contributing equally to this paper. MCS, RGT, SCOG, and RSK acknowledge financial support from the German Research Foundation (DFG) via the collaborative research center (SFB 881, Project-ID 138713538) ``The Milky Way System'' (subprojects A1, B1, B2, and B8). CDB and HPH gratefully acknowledges support for this work from the National Science Foundation under Grant No. (1816715). PCC acknowledges support from the Science and Technology Facilities Council (under grant ST/K00926/1) and StarFormMapper, a project that has received funding from the European Union's Horizon 2020 Research and Innovation Programme, under grant agreement no. 687528. HPH thanks the LSSTC Data Science Fellowship Program, which is funded by LSSTC, NSF Cybertraining Grant No. 1829740, the Brinson Foundation, and the Moore Foundation; his participation in the program has benefited this work. RJS gratefully acknowledges an STFC Ernest Rutherford fellowship (grant ST/N00485X/1) and HPC from the Durham DiRAC supercomputing facility (grants ST/P002293/1, ST/R002371/1, ST/S002502/1, and ST/R000832/1). The authors acknowledge support by the state of Baden-W\"urttemberg through bwHPC and the German Research Foundation (DFG) through grant INST 35/1134-1 FUGG. The authors gratefully acknowledge the data storage service SDS@hd supported by the Ministry of Science, Research and the Arts Baden-W\"urttemberg (MWK) and the German Research Foundation (DFG) through grant INST 35/1314-1 FUGG. 

\section*{Data availability} The data underlying this article will be shared on reasonable request to the corresponding author. Movies of the simulations can be found at the following link: \url{https://www.youtube.com/channel/UCwnzfO-xLxzRDz9XsexfPoQ}.

\def\aap{A\&A}\def\aj{AJ}\def\apj{ApJ}\def\mnras{MNRAS}\def\araa{ARA\&A}\def\aapr{Astronomy \&
 Astrophysics Review}\def\apjs{ApJS}\def\apjl{ApJ}\def\pasj{PASJ}\def\nat{Nature}\def\prd{Phys. Rev. D}
\def\ssr{Space Sci. Rev.}\def\pasp{PASP}\def\aaps{A\&AS}\def\pasa{Publications of the Astronomical Society of Australia}\def\rmp{Rev. Mod. Phys.}\def\apss{Astrophysics and Space Science}\def\baas{Bulletin of the American Astronomical Society}
\bibliographystyle{mnras}
\bibliography{bibliography}

\appendix

\section{Resolution study} \label{sec:resstudy}

In this appendix we show the results of a resolution study that we have conducted in order to assess the impact of varying the resolution and the sink particle creation threshold $\rho_{\rm c}$ (see Section~2.4 of Paper I). We consider 4 simulations, whose properties are summarised in Table \ref{tab:res}. 

\begin{figure}
	\includegraphics[width=\columnwidth]{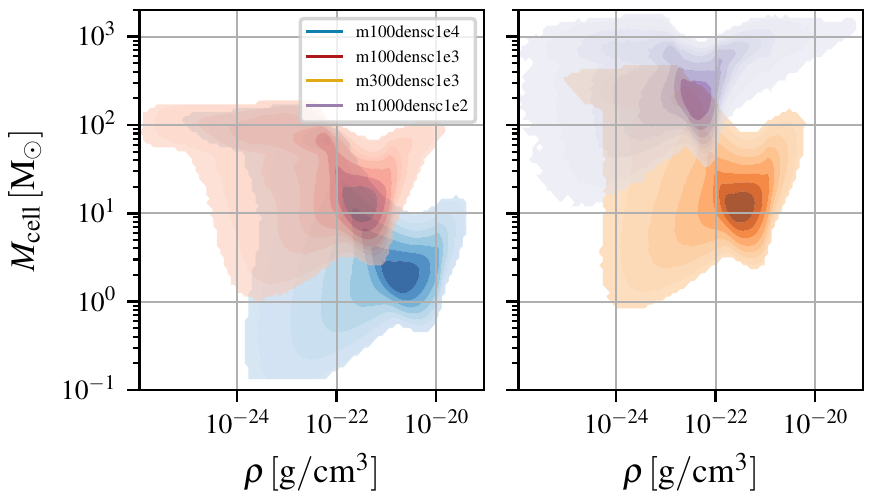}
    \caption{Mass resolution of the four simulation considered in the resolution study. The histogram shows the distribution of number of cells in the $(M_{\rm cell}, \rho)$ plane.}
    \label{fig:res_study_1}
\end{figure}

The simulations differ for two parameters: the base target cell mass $M$, and the sink particle formation density threshold $\rho_{\rm c}$. The fiducial simulation (m100densc1e4) has both the smallest $M$ (highest resolution) and the highest $\rho_{\rm c}$. The simulation m100densc1e3 has the same $M$ but lower $\rho_{\rm c}$. This allows us to assess the impact of having lower resolution in the high density regions where the gravitational collapse is happening, which is important in the context of star formation. Then we consider a simulation with a higher $M$ (i.e., lower resolution), m300densc1e3, in order to assess how a different base mass resolution affects the various phases of the ISM. Finally, we consider a very low resolution simulation, m1000densc1e2, as a general benchmark. Figure \ref{fig:res_study_1} shows the mass resolution of the four simulations as a function of density.

\begin{figure}
	\includegraphics[width=\columnwidth]{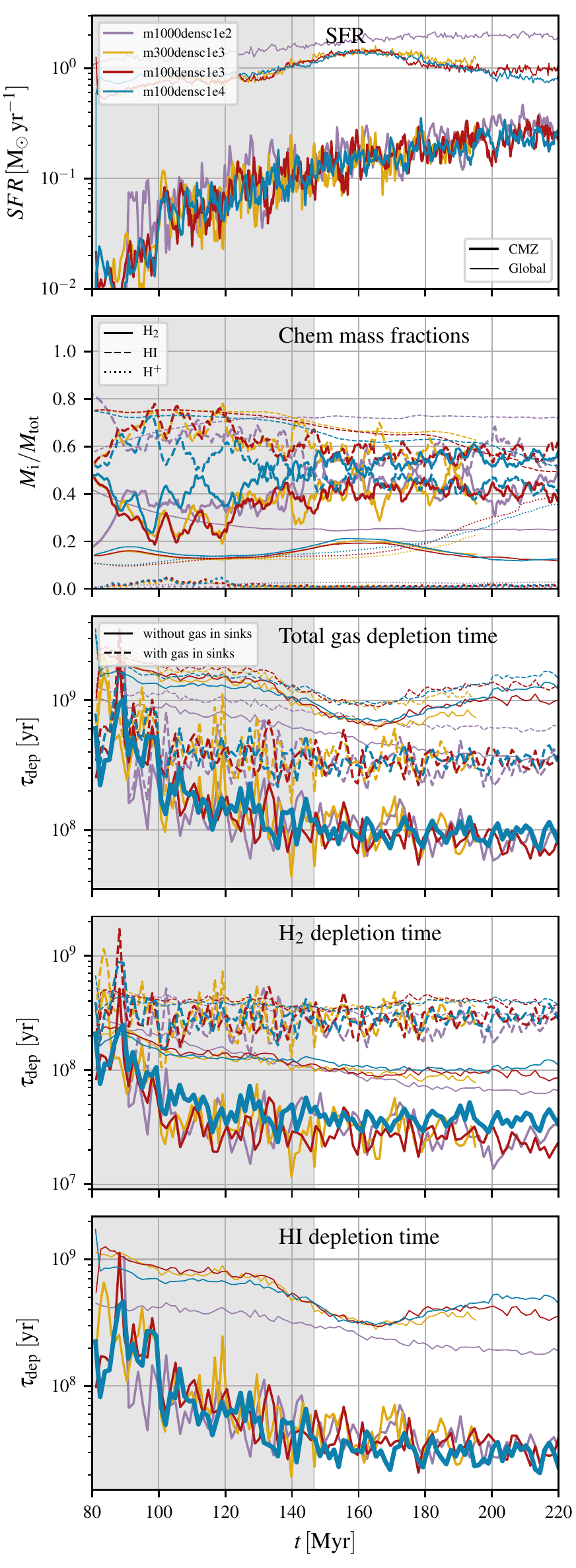}
    \caption{Various quantities as a function of time for the four simulations considered in our resolution study. Different colours indicate different simulation (see Table \ref{tab:res}). Thick lines indicate the CMZ, defined as the region within $R \leq 250 \pc$, while thin lines indicate all the simulated box.}
    \label{fig:res_study_2}
\end{figure}

Figure \ref{fig:res_study_2} shows the behaviour of various quantities as a function of time for the four simulations considered in the resolution study. From this figure we see that the largest difference between the different simulations is seen in the chemical mass fraction: at higher resolution there is roughly a factor of 2 more gas in molecular form (H$_2$) than in lower resolution simulations (see second panel from top to bottom). This induces a similar difference in the H$_2$ depletion times. The SFR and the total gas depletion times does not appear to change substantially between the different simulations. While this is encouraging and gives us confidence in the results of our main simulation, we caution against drawing too many conclusions about convergence from this. We cannot rule out that a further increase in resolution may show major differences, since the star formation process is not resolved in our simulations.

\begin{table}
	\centering
	\begin{tabular}{lcc} 
		name					& $M_{\rm base} \,\,\, [\Msun]$	& $\rho_{\rm c} \,\,\, [{\rm g\, cm^{-3}} ]$		\\
		\hline
		m1000densc1e2    			& $1000$ 					& $10^{-22}$ 						\\
		m300densc1e3 			& $300$    				& $10^{-21}$	 					\\
		m100densc1e3 			& $100$       				& $10^{-21}$ 						\\
		m100densc1e4 (fiducial)		& $100$		 			& $10^{-20}$ 						\\
		\hline
	\end{tabular}
        \caption{Summary of the simulations considered in the resolution study. $M_{\rm base}$ is the base target cell mass. No cells in the simulations are allowed to fall below this resolution (i.e., no cells can have mass higher than $M_{\rm base}$). $\rho_{\rm c}$ is the sink particle formation threshold (see Section~2.4 of Paper I). m100densc1e4 is the fiducial simulation considered for analysis in the main text of this paper and of Paper I.}
\label{tab:res}
\end{table}

\end{document}